\documentclass[journal]{IEEEtran}
\ifCLASSINFOpdf
\else
\fi

\usepackage{amsmath,amssymb,amsfonts}
\usepackage{algorithmic}
\usepackage{graphicx}
\usepackage{textcomp}			
\usepackage{amsthm}
\usepackage{amsfonts} 			
\usepackage{bm} 
\usepackage{float}
\usepackage[colorlinks,linkcolor=blue,anchorcolor=blue,citecolor=green,CJKbookmarks=True]{hyperref}	
\usepackage{multirow}
\usepackage{booktabs}
\usepackage{subfig}
\usepackage{caption}
\usepackage{xcolor}
\usepackage{algorithm}
\usepackage{xspace}
\usepackage{overpic}

\newcommand*{\ie}{\textit{i.e.}\@\xspace}
\newcommand*{\etal}{\textit{et al.}\@\xspace}
\theoremstyle{plain}
\newtheorem{thm}{Theorem}

\begin{document}
%
\title{Mixed Noise Removal with Pareto Prior}
%
%
%

\author{Zhou~Liu,
        Lei~Yu,~\IEEEmembership{Member,~IEEE,}
        Gui-Song~Xia,~\IEEEmembership{Senior Member,~IEEE}
        and~Hong~Sun,~\IEEEmembership{Member,~IEEE}
\thanks{	Zhou Liu and Gui-Song Xia are with School of Computer Science, Wuhan University, Wuhan, China. 
	
	Lei Yu and Hong Sun are with the School of Electronic and Information, Wuhan University, Wuhan, China (e-mail: ly.wd@whu.edu.cn)
}. 
}

%
%

\markboth{Journal of \LaTeX\ Class Files,~Vol.~14, No.~8, August~2015}%
{Shell \MakeLowercase{\textit{et al.}}: Bare Demo of IEEEtran.cls for IEEE Journals}
%



\maketitle

\begin{abstract}

Denoising images contaminated by the mixture of additive white Gaussian noise (AWGN) and impulse noise (IN) is an essential but challenging problem. The presence of impulsive disturbances inevitably affects the distribution of noises and thus largely degrades the performance of traditional AWGN denoisers.  Existing methods target to compensate the effects of IN by introducing a weighting matrix, which, however, is lack of proper priori and thus hard to be accurately estimated. To address this problem, we exploit the Pareto distribution as the priori of the weighting matrix, based on which an accurate and robust weight estimator is proposed for mixed noise removal. Particularly, a relatively small portion of pixels are assumed to be contaminated with IN, which should have weights with small values and then be penalized out. This phenomenon can be properly described by the Pareto distribution of type 1. Therefore, armed with the Pareto distribution, we formulate the problem of mixed noise removal in the Bayesian framework, where nonlocal self-similarity priori is further exploited by adopting nonlocal low rank approximation. Compared to existing methods, the proposed method can estimate the weighting matrix adaptively, accurately, and robust for different level of noises, thus can boost the denoising performance. Experimental results on widely used image datasets demonstrate the superiority of our proposed method to the state-of-the-arts.

\end{abstract}


%
\IEEEpeerreviewmaketitle

\section{Introduction} \label{Sec:intro}

Removing noise from images is a longstanding but important problem in image processing. A variety of image denoising methods have been proposed with Gaussian noise assumption \cite{elad2006image,buades2005non,dabov2007image, mairal2009non,dong2013nonlocal,gu2016weighted,liu2018image}. However, due to the faulty memory locations in hardware or bit errors in transmission, images can be further contaminated by impulse noise (IN) \cite{yan2013restoration}. Generally, there are two typical types of IN, \textit{i.e.}, salt-and-pepper impulse noise (SPIN) and random-valued impulse noise (RVIN) \cite{dong2007detection}. And the presence of IN inevitably affects the distribution of noise and largely degrade the performance of traditional AWGN denoisers \cite{elad2006image,buades2005non,dabov2007image, gu2016weighted}.  Therefore, removing mixed noise (mixture of AWGN and IN) in image is an essential but challenging problem \cite{lopez2010restoration}.

\begin{figure}[t]\setcounter{subfigure}{0} 
	\centering
	\setlength{\abovecaptionskip}{.3cm}
	\setlength{\belowcaptionskip}{-.3cm}

		\subfloat{
		\begin{minipage}{0.165\textwidth}
			\centering
			\begin{overpic}[width=1\textwidth]{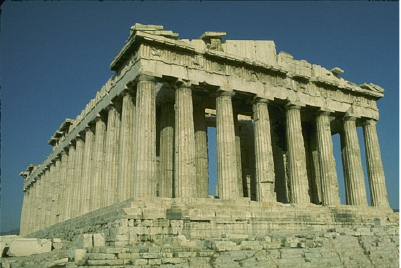}
      	    \put(35,-8){\scriptsize\bfseries{(a) Original}}
          \end{overpic}
		\end{minipage}
	}  	\hspace*{-0.3cm}
	\subfloat{
		\begin{minipage}{0.165\textwidth}
			\centering
			\begin{overpic}[width=1\textwidth]{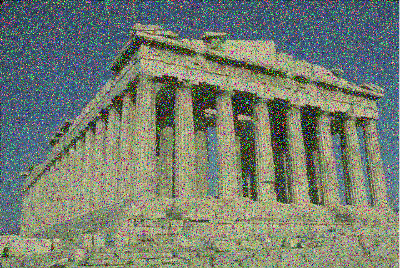}
				  \put(35,-8){\scriptsize\bfseries{(b) Noisy}}
            \end{overpic}
		\end{minipage}
	}  		\hspace*{-0.3cm}
	\subfloat{
		\begin{minipage}{0.165\textwidth}
			\centering
			\begin{overpic}[width=1\textwidth]{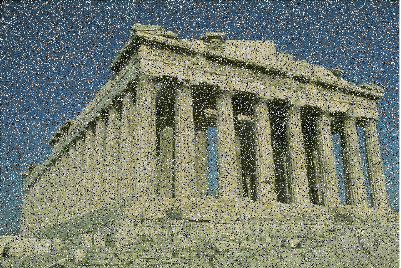}
	\put(25,-8){\scriptsize\bfseries{(c) WJSR \cite{liu2017weighted}}}
		\put(25,60){\scriptsize\bfseries\color{yellow}{PSNR: 14.33}}
\end{overpic}
		\end{minipage}
	}  	\\ 	
\subfloat{
	\begin{minipage}{0.165\textwidth}
		\centering
			\begin{overpic}[width=1\textwidth]{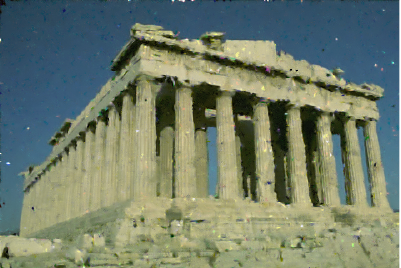}
	\put(25,-8){\scriptsize\bfseries{(d) WESNR \cite{jiang2014mixed}}}
	\put(25,60){\scriptsize\bfseries\color{yellow}{PSNR: 24.34}}
\end{overpic}

	\end{minipage}
}  		\hspace*{-0.3cm}
	\subfloat{
		\begin{minipage}{0.165\textwidth}
			\centering
    	\begin{overpic}[width=1\textwidth]{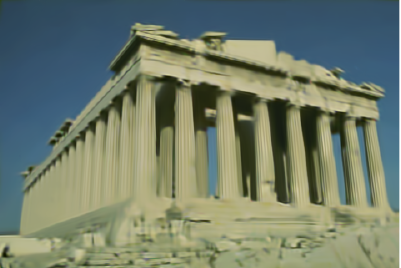}
      	\put(25,-8){\scriptsize\bfseries{(e) RCSR \cite{liu2018mixed}}}
    	\put(25,60){\scriptsize\bfseries\color{yellow}{PSNR: 26.06}}
      \end{overpic}
		\end{minipage}
	}  		\hspace*{-0.3cm}
	\subfloat{
		\begin{minipage}{0.165\textwidth}
			\centering
    	\begin{overpic}[width=1\textwidth]{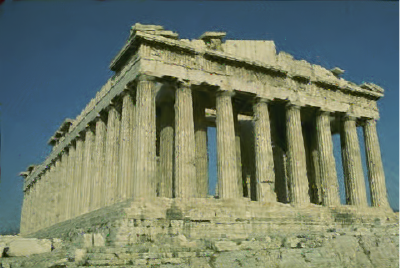}
    	\put(16,-8){\scriptsize\bfseries{(f) Our RWE method}}
	  \put(25,60){\scriptsize\bfseries\color{yellow}{PSNR: 28.07}}
      \end{overpic}
		\end{minipage}
	}  	


%
\subfloat{
	\begin{minipage}{0.49\textwidth}
		\centering
		\begin{overpic}[width=1\textwidth]{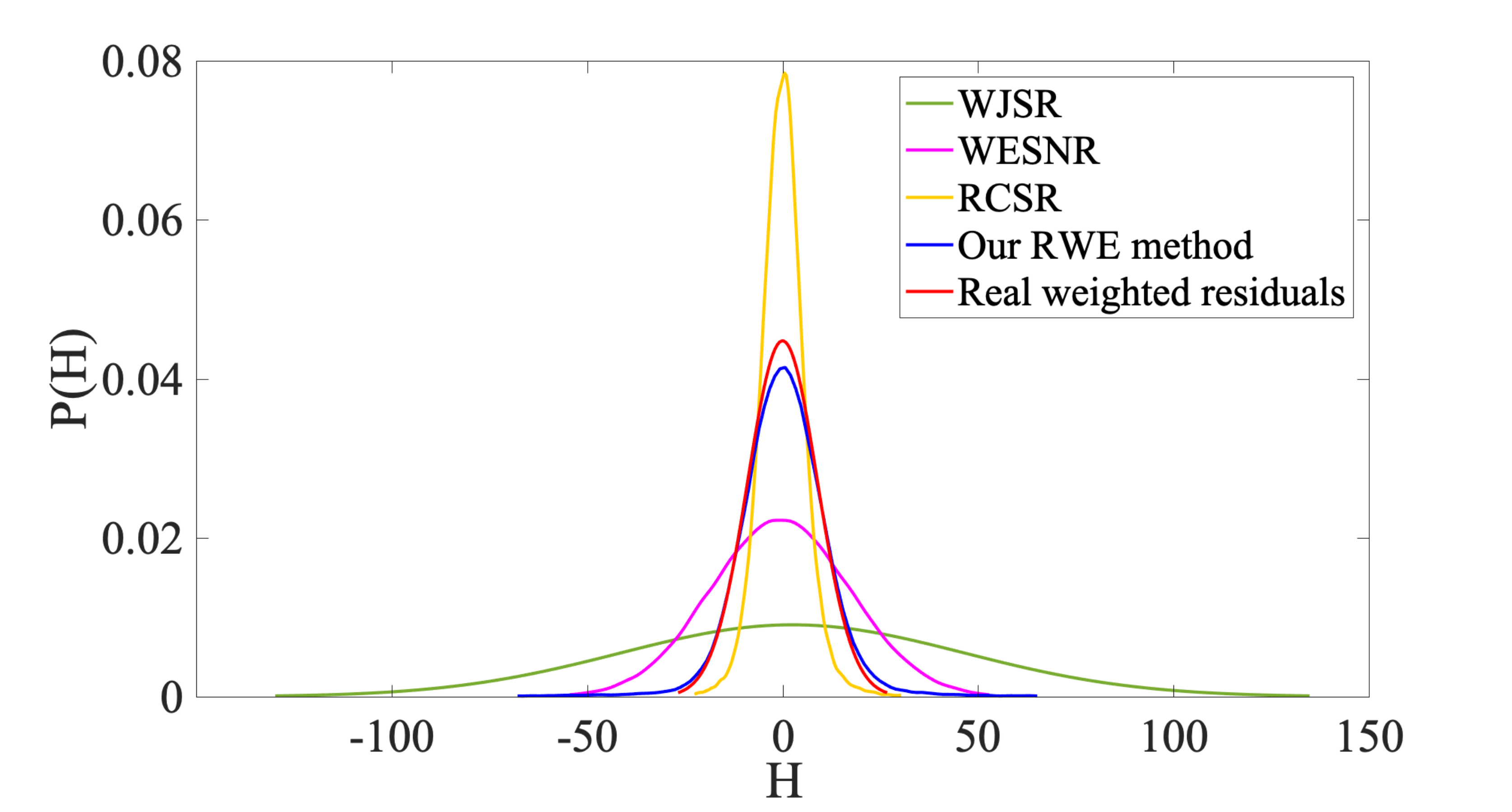}
			\put(50,-3){\scriptsize\bfseries{(g)}}
		\end{overpic}
	\end{minipage}
}  	\\

\caption{Denoising of image corrupted by mixed noise: (a) Original image, (b) Noisy image (corrupted with AWGN+RVIN+SPIN), (c)-(f) are recovered images by: WJSR \cite{liu2017weighted},  WESNR \cite{jiang2014mixed}, RCSR\protect\footnotemark[1] \cite{liu2018mixed}, our RWE method. (g) is the corresponding histograms of weighted residuals obtained in (c)-(f). Apparently, our RWE method can estimate the weighting matrix more accurately and thus achieve state-of-the-art performance.}
	\label{fig:highlight}
\end{figure}
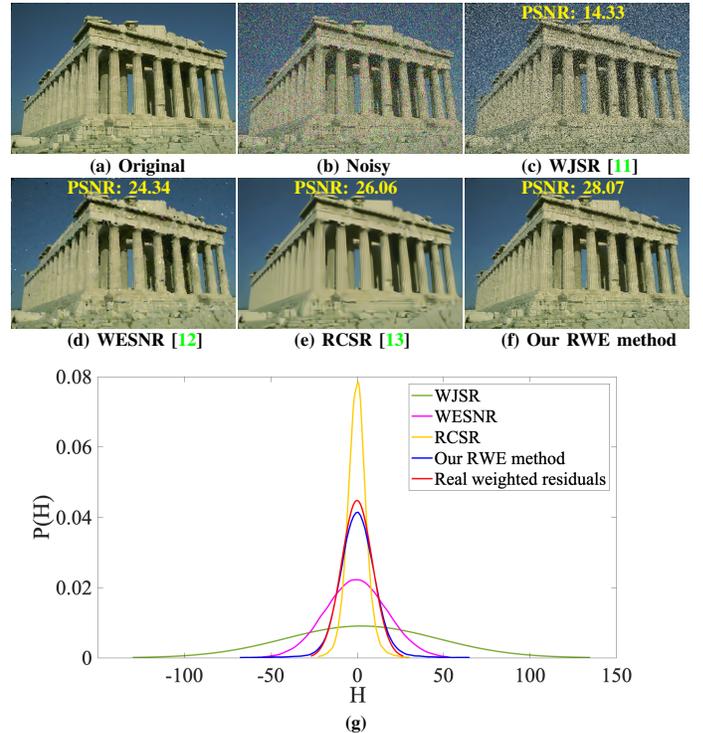



Given an image $\bm{Y}$ contaminated by mixed noise, to recover a noise-free image $\bm{X}$, several methods \cite{liu2017weighted,jiang2014mixed,liu2018mixed,xiao2011restoration,jiang2015mixed} introduce a weighting matrix $\bm{W}$ to compensate the effects of IN.  By modeling the weighted residuals $\bm{H} = \bm{W}\odot(\bm{Y}-\bm{X})$ approximately with Gaussian distribution, where $\odot$ denotes the element-wise product, the AWGN denoisers are feasible for mixed noise removal. Evidently, the weighting matrix plays an important role in these methods. When the weighting matrix is over-estimated, the effects of IN cannot be suppressed effectively, as shown in Fig.~\ref{fig:highlight}(d). While the weighting matrix is under-estimated, the result tends to over-smooth the fine details in image, as shown in Fig.~\ref{fig:highlight}(e). Hence the key issue of these methods is to accurately estimate the weighting matrix.

\footnotetext[1]{The codes of RCSR is not released and we integrate the prior proposed by RCSR into our model to implement it.} 

The existing methods to estimate the weighting matrix can be roughly divided into two categories: \textit{Hard IN detection based methods} \cite{liu2017weighted,xiao2011restoration} and \textit{Soft IN detection based methods} \cite{jiang2014mixed,liu2018mixed,jiang2015mixed}. Hard IN detection based methods  \cite{liu2017weighted,xiao2011restoration} binarily compensate the effects of IN by directly resetting the IN corrupted pixels which indicated by IN detectors. Nevertheless, the denoising performance largely relies on the accuracy of IN detectors. Moreover, it is not easy to apply this kind of methods to handle the mixture of AWGN, SPIN and RVIN, as shown in Fig.~\ref{fig:highlight} (c). Soft IN detection based methods \cite{jiang2014mixed,liu2018mixed,jiang2015mixed} attempt to compensate the effects of IN without IN detectors, instead, the weighting matrix is adaptively estimated. However, they still lack of proper prior to accurately and robustly estimate weighting matrix under various mixed noise environments.

Therefore, the proper prior is crucial for the accurate weighting matrix estimation. In this work, we target to design a proper prior of the weighting matrix for mixed noise removal. For images contaminated by mixed noise, it can be observed that a relatively small portion of pixels are assumed to be contaminated with IN, which should be assigned with small weights, and the remaining pixels are assumed to be corrupted with AWGN, which should have large weights. Thus, the behavior of the inverse of weighting matrix can be described by the Pareto distribution of type 1. Based this observations, we propose a mixed noise removal method with Pareto prior. Specifically, armed with the Pareto prior, we formulate the problem of mixed noise removal in the Bayesian framework. To fully exploit nonlocal self-similarity, we adopt a non-local low rank approximation term as the image prior. After implementing Bayesian inference, our adaptive and robust weight estimation model is obtained. A splitting Bregman \cite{goldstein2009split} based alternative optimization algorithm is developed to solve our proposed model. Extensive experiments are conducted to demonstrate our proposed method can achieve state-of-the-art performance.

Our main contributions are summarized as follows,
\begin{itemize}
%
%

\item We propose to employ the Pareto distribution to describe the behavior of the inverse of weighting matrix. This makes the distribution of weighs coincide with the probability of pixels uncorrupted by IN, on which bases we can estimate the weighting matrix accurately and robustly for various levels of mixed noise.

\item We present a novel Robust Weight Estimation (RWE) model for mixed noise removal, where the Pareto prior and nonlocal self-similarity are exploited together. 
	
	
\item Experimental results on commonly used image datasets verify the superior performance of our proposed RWE method, in terms of both quantitative and qualitative results. 

	
\end{itemize}

The rest of this paper is organized as follows: we review the most related work in Section~\ref{Sec:related}. In Section \ref{Sec:Proposed}, we describe in detail our proposed mixed noise removal method. The optimization and analysis for our proposed model are presented in Section \ref{Sec:Optimization}. Section \ref{Sec:Experiment} shows some experimental results and Section \ref{Sec:Conclusion} reaches a conclusion.

%
%
%



\section{Related Works} \label{Sec:related}

A variety of mixed noise removal methods have been proposed.  We briefly review the classic and contemporary works closely related to ours.

Intuitively, for an image corrupted by mixed noise, the IN corrupted pixels can be detected and replaced with some statistics (\textit{e.g.}, median), then remove the AWGN in the remaining pixels. This two-stage strategy (detecting then filtering strategy) has been adopted in \cite{garnett2005universal, cai2008two, xiong2012universal} for mixed noise removal. However, such two-stage strategy usually become less efficient and generate many artifacts when the level of mixed noise is high \cite{jiang2014mixed}. In order to remove IN and AWGN simultaneously, the statistical distribution of mixed noise is attempted to characterize in \cite{lopez2010restoration,zhuang2016mixed,huang2017mixed} such that the noise statistics can be adaptively inferred. The mixture of Gaussian and triangular distribution was employed in \cite{lopez2010restoration} to model the denoised errors. To avoid empirical parameters, \cite{zhuang2016mixed} proposed to impose the spike-slab prior for IN and infer noise statistics in a non-parametric manner. To directly model the distribution of IN, LSM-NLR \cite{huang2017mixed} was proposed to utilize LSM (Laplacian Scale Mixture) distributions. However, it is not easy to characterize the distribution of mixed noise by a parametric model. And the complicated distribution often makes the mixed noise removal troublesome.

In order to take advantages of traditional AWGN denoisers, several methods \cite{liu2017weighted,jiang2014mixed,liu2018mixed,xiao2011restoration,jiang2015mixed} introduce a weighting matrix embedded into the data fidelity term to suppress the effects of IN and makes the weighted residuals close to Gaussian distribution. Thus the key point of these methods is to accurately estimate the weighting matrix. Hard IN detection based methods \cite{liu2017weighted,xiao2011restoration} attempt to obtain binary weighting matrix by using IN detectors, where a representative work is WJSR \cite{liu2017weighted}. Nevertheless, the denoising performance largely relies on the accuracy of IN detectors. Moreover, it is not easy to apply this kind of methods to handle the mixture of AWGN, SPIN and RVIN. To avoid the explicit IN detection, soft IN detection based methods \cite{jiang2014mixed,liu2018mixed,jiang2015mixed} are proposed to estimate the weighting matrix adaptively.  To achieve this end, WESNR \cite{jiang2014mixed} was proposed to estimate the weighting matrix from the encoding residual using a Gaussian kernel function. But the decreasing rate of weights is controlled by a empirical parameter, which is very sensitive to the mixed noise level, especially for AWGN+SPIN+RVIN. To address this problem, the priori of weighting matrix is exploited to estimate the weighting matrix adaptively and robustly. RCSR \cite{liu2018mixed} proposed an entropy-like priori and achieved a robust method for mixed noise removal, which in some sense alleviate the problem. However,  the entropy priori is not proper and often exacerbates the overweighting problem, thus leading to over-smoothing result. Therefore, the accurate weighting matrix estimation is very challenging due to the lack of proper prior. In this work, we target to design a proper prior of the weighting matrix for mixed noise removal.

To learn the valid prior from the external datasets, deep learning based methods \cite{wang2019variational,ding2019improved} have been proposed for mixed noise removal. The merge of deep convolutional neural networks and variational model \cite{wang2019variational} has boosted the performance of mixed noise removal. A CNN process as a regularization term was integrated in the variational model \cite{liu2013weighted}. However, these methods need to train from a large mount of labeled samples and just work for a specific type of mixed noise (AWGN+SPIN or AWGN+RVIN). In this work, we aim to flexibly handle three types of mixed noise and only need single image.

 \begin{figure}[h]
	\centering 
	\setlength{\abovecaptionskip}{0cm}
	\setlength{\belowcaptionskip}{-0.3cm}
	\includegraphics[width=0.8\columnwidth]{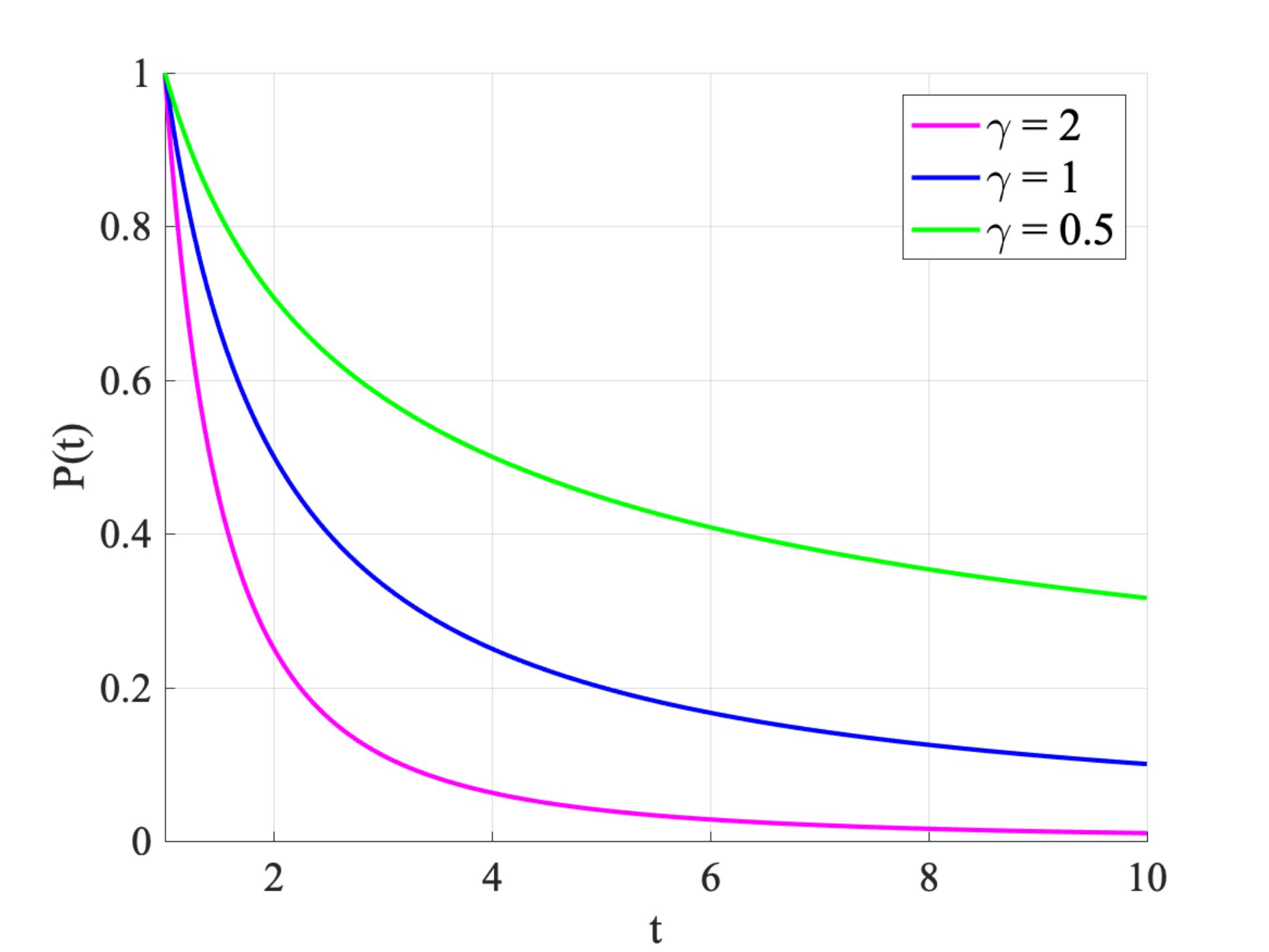}                                    
	\caption{Pareto Type 1 distribution with unit scale parameter and vairous shape parameter $\gamma$.}
	\label{distribution}
\end{figure}

\section{ Mixed Noise Removal with Pareto Prior}  \label{Sec:Proposed}




%


\subsection{Mixed Noise Model}

Considering noise-free image $\bm{X}\in  \mathbb{R}^{M \times N}$, observation $\bm{Y}\in \mathbb{R}^{M\times N}$ corrupted by mixed noise could be written as follows:
\begin{equation} \label{noise}
\bm{Y}_{i,j} = \left \{ 
\begin{aligned}
&\bm{X}_{i,j} + \bm{G}_{i,j} \qquad  & {i,j} \notin \mathcal{I}
\\
&\bm{I}_{i,j}   \qquad &{i,j} \in \mathcal{I}
\end{aligned} 
\right.
\end{equation}
where the subscript ${i,j}$ denotes the pixel of image at location $(i,j)$, $\bm{G}_{i,j}$ represents the AWGN at location $(i,j)$ and $\mathcal{I}$ denotes the set of pixels corrupted by IN. 

%

Considering the two typical types of IN, \ie, SPIN and RVIN, Eq.~\eqref{noise} can be one of the following three forms that corresponding to three different types of mixed noises:
%


\begin{itemize}
	\item \textbf{AWGN+SPIN}: 
	\begin{equation} \label{awgnsp}
	\bm{Y}_{i,j} = \left \{ 
	\begin{aligned}
	&\bm{X}_{i,j} + \bm{G}_{i,j}    &\quad   \text{with probability}  \quad &1-s 
	\\
	&v_{min}   \quad    &\text{with probability}  \quad  &s/2
	\\
	&v_{max} \quad    &\text{with probability} \quad  &s/2
	\end{aligned}   
	\right.
	\end{equation}
	where $[v_{min}, v_{max}]$ denotes the dynamic range of images, $ 0 \le s \le 1$ indicates the ratio of pixels that contaminated by SPIN. An exemplar corrupted by AWGN+SPIN can be found in Fig.~\ref{Exp:lena}. 
	
	\item \textbf{AWGN+RVIN}:
	\begin{equation} \label{awgnrv}
	\bm{Y}_{i,j} = \left \{ 
	\begin{aligned}
	&\bm{X}_{i,j} + \bm{G}_{i,j}    &\quad   \text{with probability}  \quad &1-r 
	\\
	&q   \quad    &\text{with probability}  \quad  &r
	\end{aligned}   
	\right.
	\end{equation}
	where $ 0 \le r \le 1$ and $q$ is a random value uniformly distributed in $[v_{min}, v_{max}]$. An exemplar corrupted by AWGN+RVIN can be found in Fig.~\ref{Exp:barbara}. 
	
	\item \textbf{AWGN+SPIN+RVIN}:
	\begin{equation} \label{awgnsprv}
	\small 
	\bm{Y}_{i,j} = \left \{ 
	\begin{aligned}
	&\bm{X}_{i,j} + \bm{G}_{i,j}    &\quad   \text{with probability}  \quad &(1-s)(1-r) 
	\\
	&v_{min}   \quad    &\text{with probability}  \quad  &s/2
	\\
	&v_{max} \quad    &\text{with probability} \quad  &s/2
	\\
	&q   \quad    &\text{with probability}  \quad  &r(1-s)
	\end{aligned}   
	\right.
	\end{equation}
An exemplar corrupted by AWGN+RVIN can be found in Fig.~\ref{Exp:peppers}. 
\end{itemize}

For simplicity, the dynamic range of image can be normalized in $[0,1]$, \ie, $v_{min} = 0$ and $v_{max} = 1$.

\subsection{Mixed Noise Removal Model from Bayesian Perspective}

%

Based on mixed noise model Eq.~\eqref{noise}, a weighting matrix $\bm{W}$ can be introduced to suppress the effects of IN. Then the weighted residuals $\bm{H} = \bm{W}\odot(\bm{Y}-\bm{X})$  can be approximately modeled as Gaussian distribution with zero mean. Therefore, the mixed noise removal problem we encountered is to jointly estimate the noise-free image $\bm{X}$, the weighting matrix $\bm{W}$ and the variance $\sigma$ of Gaussian distribution from the noisy observation $\bm{Y}$. This problem can be formulated as the following   \textit{maximum a posteriori (MAP)} problem:


\begin{equation}\label{MAP}
\begin{aligned}
\underbrace{P(\bm{X}, \bm{W}, \sigma | \bm{Y})}_{\text{posterior}} \propto \prod_{i,j}^{M,N}\underbrace{ P\left(\bm{Y}_{i,j} |\bm{X}_{i,j}, \bm{W}_{i,j}, \sigma \right)}_{\text { likelihood }} \underbrace{ P\left(\bm{X}_{i,j}  \right)}_{\text { prior on $\bm{X}$ }} \\  \underbrace{ P\left(\bm{W}_{i,j}  \right)}_{\text { prior on $\bm{W}$}} \underbrace{ P\left(\sigma \right)}_{\text { prior on $\sigma$}} 
\end{aligned}
\end{equation}


The likelihood of the weighted residuals is characterized by Gaussian distribution, \ie,
\[
P\left(\bm{Y}_{i,j} |\bm{X}_{i,j}, \bm{W}_{i,j}, \sigma \right) = \frac{1}{\sqrt{2\pi} \sigma} \exp \left(-\frac{\bm{W}_{i,j}^2(\bm{Y}_{i,j}-\bm{X}_{i,j})^2}{2\sigma^2}\right) 
\]
where the variance  $\sigma$ is assumed to follow an improper prior \cite{gelman2013bayesian},
\[
P(\sigma) \propto 1.
\]
The priors of image $\bm{X}$ and weights $\bm{W}$ are then to be designed. 


\subsection{Pareto Type 1 Distribution and Weighting Matrix Prior}

Pareto distribution is originally applied to describing the distribution of wealth in a society \cite{reed2003pareto}, then many situations in which an equilibrium is found in the distribution of the "small" to the "large", such as file size of Internet traffic \cite{reed2004double}, hard disk error rates \cite{schroeder2010understanding}, \etal  Analogically, in mixed noise removal problem, it can be observed that a relatively small portion of pixels are assumed to be contaminated with IN, which should be assigned with small weights, and the remaining pixels are assumed to be corrupted with AWGN, which should have large weights (close to one). Hence we have the fact that few IN (the inverse of weights are large) but many AWGN contaminated pixels (the inverse of weights are small). Consequently, it is nature to apply Pareto type 1 distribution with unit scale parameter to describing the behavior of the inverse of weighting matrix,

\begin{figure}[t]\setcounter{subfigure}{0} 
	\centering
	\setlength{\abovecaptionskip}{0cm}
	\setlength{\belowcaptionskip}{-0.5cm}
	\subfloat{
		\begin{minipage}{0.4\textwidth}
			\centering
			\includegraphics[height=0.8\columnwidth]{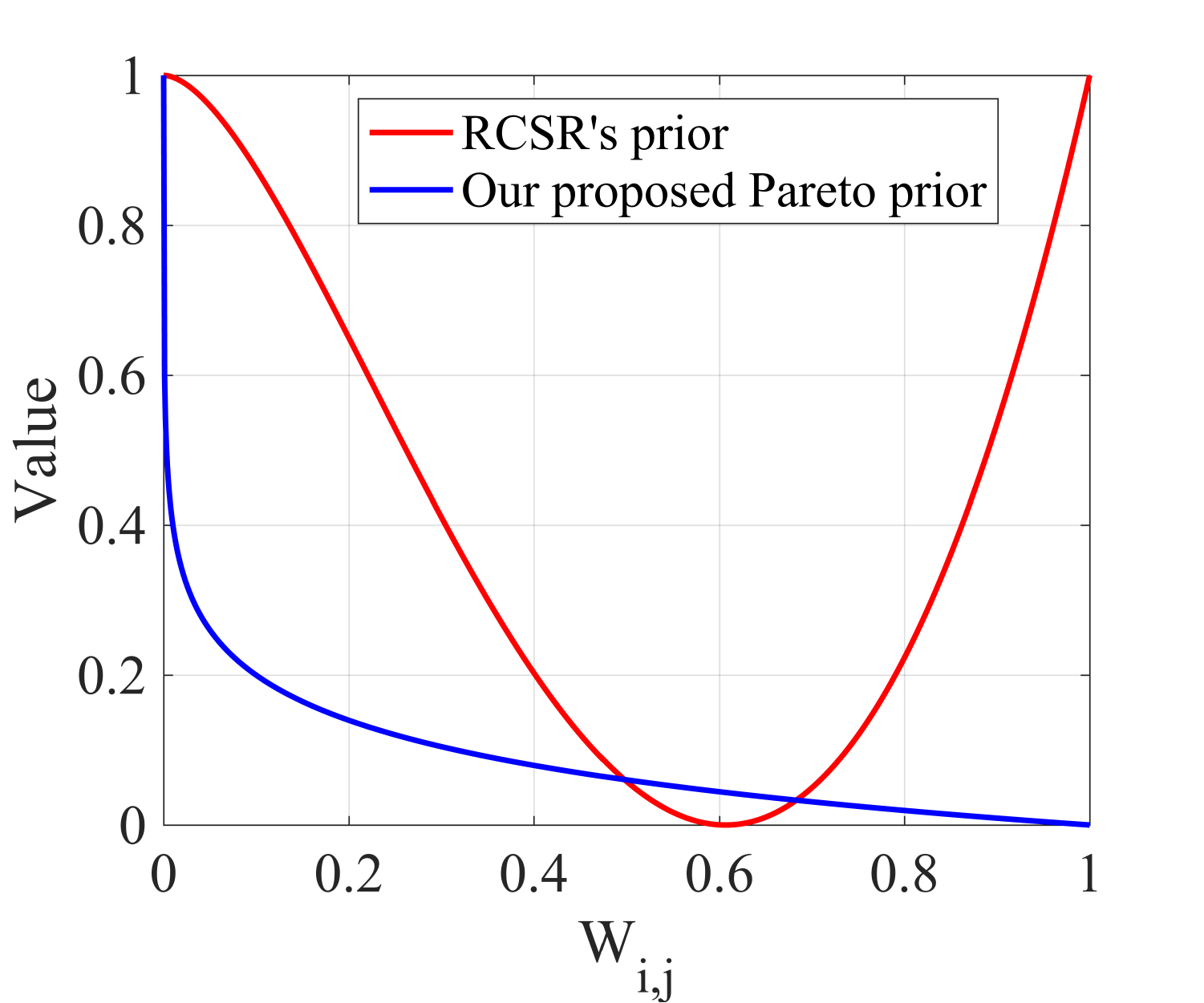}
			\label{curve:a}                                       
		\end{minipage}
	}  	
	\caption{The curve of RCSR's prior (\ref{RCSRreg}) (\textcolor[RGB]{255,0,0}{Red}), our proposed prior (\ref{RWEreg}) (\textcolor[RGB]{0,0,255}{Blue}).} 
	\label{curve}
\end{figure}

\begin{equation}\label{weighting_matrix}
	\bm{W}_{i,j} \sim \left(\frac{1}{\bm{W}_{i,j}}\right)^{-\bm{\gamma}_{i,j}} 
\end{equation}
where $\bm{\gamma}_{i,j}>0$ is the shape parameter. 
%
We plot the probability density function of Pareto Type 1 distribution in Fig.~\ref{distribution}. Obviously, larger $\bm{\gamma}_{i,j}$ leads to $\bm{W}_{i,j}$ approaching 1 corresponding to small portion of IN corrupted pixels.

And by adjusting $\bm{\gamma}_{i,j}\to 0$, the limit distribution will be uniform for $\bm{W}_{i,j}$ in the range of $[0,1]$ and thus become a non-informative priori. Hence, we have the following Pareto prior
\begin{equation}\label{weightprior}
P(\bm{W}_{i,j}) \propto \bm{W}_{i,j}^{\;\bm{\gamma}_{i,j}}
\end{equation}
We will introduce the adaptive setting of $\bm{\gamma}_{i,j}$ in the Section \ref{Subsec:gamma}.
 
\subsection{Low Rank Matrix Approximation}
As natural images often have many repetitive patterns, \ie, a patch can have many similar patches across the whole image, the nonlocal self-similarity has been proven to be a very effective prior for image denoising \cite{buades2005non,dabov2007image,dong2013nonlocal,gu2016weighted}.  To fully exploit nonlocal self-similarity, low rank matrix approximation (LRMA) has been applied in image denoising with significant improvements \cite{dong2013nonlocal,gu2016weighted}. Thus, in this work, we adopt the low rank approximation to exploit the nonlocal self-similarity. 

For each exemplar vectorized patch $\bm{x}_{p} \in \mathbb{R}^{m}$, the sets of similar patches can be formed by collecting the first $K$ most similar patches to $\bm{x}_{p}$, denoted as $\bm{R}_{p}\bm{X} \in \mathbb{R}^{m \times K}$, where $\bm{R}_{p}$ denotes the operator extracting the similar patches. 
Assuming that each group of similar patches are independent, the patch based nonlocal low rank image prior can be formulated as 
\begin{equation} \label{lowrank}
\begin{aligned}
&P(\bm{X}) \propto \prod_{p=1}^{C} P(\bm{R}_{p}\bm{X})\\
&P(\bm{R}_{p}\bm{X}) \propto \exp(-\lambda \mbox{Rank}(\bm{R}_{p}\bm{X}))
\end{aligned}
\end{equation}
where $C$ is the total numbers of exemplar patches, $\mbox{Rank}(\cdot)$ is the function to calculate the rank of matrix and $\lambda$ is the balance parameter. 

\subsection{The Full Proposed Model}
By substituting (\ref{weightprior}) and (\ref{lowrank}) into (\ref{MAP}), maximizing the posterior  (\ref{MAP}) is equivalent to minimizing its negative logarithm, 
\begin{equation} \label{Propose1}
\begin{aligned}
\min_{\bm{X},\bm{W}, \sigma} \sum_{i,j}^{M,N}\frac{1}{2 \sigma^{2}} &\left( \bm{W}_{i,j} ( \bm{Y}_{i,j}-\bm{X}_{i,j}) \right)^{2}- \sum_{i,j}^{M,N}  \bm{\gamma}_{i,j}\log(\bm{W}_{i,j}+\epsilon)\\
&+\lambda \sum_{p=1}^{C} \mbox{Rank}(\bm{R}_{p}\bm{X})+ MN\log\sigma
\end{aligned}
\end{equation}
where $\epsilon$ is a very small constant for numerical stability and we fix it at $10^{-6}$. 

Since minimizing $\mbox{Rank}(\cdot)$ is a NP-hard problem, the nuclear norm $\left\| \cdot\right\|_{*}$ is widely used for the convex relaxation \cite{candes2010power}. And considering that $\bm{W}_{i,j}$ is in $[0,1]$, we can reformulate \eqref{Propose1} as 

\begin{equation}\label{Propose2}
\begin{aligned}
\min_{\bm{X},\bm{W}, \sigma} \frac{1}{2\sigma^{2}} &\left\| \bm{W} \odot (\bm{Y}-\bm{X})  \right\| _{F}^{2}+\lambda \sum_{p=1}^{C} \left\| \bm{R}_{p}\bm{X}\right\|_{*} \\ 
&+  \sum_{i,j}^{M,N}\vert \bm{\gamma}_{i,j}\log(\bm{W}_{i,j}+\epsilon)\vert +  MN\log\sigma \qquad \\
& \qquad \qquad \qquad \mbox{s.t.}  \quad 0 \le \bm{W}_{i,j} \le 1 
\end{aligned}
\end{equation}
where the cleaned image $\bm{X}$, the weight matrix $\bm{W}$ and the variance of AWGN $\sigma$ can be estimated simultaneously from the noisy observed image $\bm{Y}$.

\subsection{Discussion} \label{Subsection:dis}

Currently, very few studies focus on investigating the prior of $\bm{W}$. To the best of the authors' knowledge, only in RCSR \cite{liu2018mixed}, the authors have been working on employing the entropy of squared weights and trying to regularize the weighting matrix by maximizing the entropy. Then $\bm{W}$ in  RCSR \cite{liu2018mixed} can be estimated by optimizing the following subproblem 
\begin{equation} \label{updatewRCSR}
\min_{\bm{W}}  \left\|\bm{W} \odot (\bm{Y}-\bm{X}) \right\| _{F}^{2}+\xi \sum_{i,j}^{M,N}\bm{W}_{i,j}^{2} \log(\bm{W}_{i,j}^{2})
\end{equation}
and the corresponding solution of (\ref{updatewRCSR}) can be obtained as 
\begin{equation}\label{updatew2}
\bm{W}_{i,j}=\exp \left(-\frac{1}{2}-\frac{(\bm{Y}_{i,j}-\bm{X}_{i,j})^{2}}{2 \xi}\right)
\end{equation}
where $\xi$ is a positive constant to control the decreasing rate of weights.

For comparison, we rewrite the prior proposed by RCSR in \eqref{updatewRCSR}
\begin{equation} \label{RCSRreg}
J_{\text{RCSR}}(\bm{W}) = \xi \sum_{i,j}^{M,N}\bm{W}_{i,j}^{2} \log(\bm{W}_{i,j}^{2})
\end{equation}
and our proposed Pareto prior, denoted by $J_{\text{RWE}}$ according to \eqref{Propose2},
\begin{equation} \label{RWEreg}
J_{\text{RWE}}(\bm{W}) = \sum_{i,j}^{M,N}  - \bm{\gamma}_{i,j}\log(\bm{W}_{i,j}+\epsilon)
\end{equation}


The curves of $J_{\text{RCSR}}(\bm{W})$ and $J_{\text{RWE}}(\bm{W})$ with respect to $\bm{W}_{i,j}\in [0,1]$ are illustrated in Fig.~\ref{curve}. Obviously, $J_{\text{RCSR}}(\bm{W})$ has one minima at $\exp(-\frac{1}{2})$. In other words,  $J_{\text{RCSR}}(\bm{W})$ encourages the value of weights close to $\exp(-\frac{1}{2}) \approx 0.6$, which is not reasonable. Besides, we can see that the weight $\bm{W}_{i,j}$ estimated by Eq.\eqref{updatew2} is always in $[0, exp(-\frac{1}{2})]$, no matter how $\xi$ is set.

On the other hand, $J_{\text{RWE}}(\bm{W})$ is monotone decreasing and takes $1$ as the minima which coincides with \textit{fewer IN and many AWGN}. Together with the monotone increasing data fidelity shown in \eqref{Propose2} and balancing parameter $\gamma_{i,j}$, the weights estimator is to find the equilibrium between the data fidelity and the Pareto prior.



Based on the above analysis, it is easy to conclude that our proposed $J_{\text{RWE}}(\bm{W}) $ following the Pareto Type 1 prior is more suitable for mixed noise removal. Experimental results in Section~\ref{Exp:effective} can illustrate the effectiveness of our proposed $J_{\text{RWE}}(\bm{W})$.

\section{Optimization and Analysis} \label{Sec:Optimization}

In this section,  we develop an alternative optimization algorithm to solve (\ref{Propose2}) and find the solution of $\bm{X}, \bm{W},\sigma$ alternatively by iterations. Optimization parameter setting and convergence analysis are presented afterwards.

\subsection{Optimization Algorithm}

\subsubsection{Solving the $\mathbf{W}$ subproblem}
With the fixed $\bm{X}$ and $\sigma$, each $\bm{W}_{i,j}$ can be independently updated by solving the following scalar minimization problem
\begin{equation}\label{Wsubproblem}
\begin{aligned}
\hat{\bm{W}}_{i,j}=\arg\min_{\bm{W}_{i,j}}& \sum_{i,j}^{M,N}\frac{1}{2 \sigma^{2}} \left( \bm{W}_{i,j} ( \bm{Y}_{i,j}-\bm{X}_{i,j}) \right)^{2} \\
&-  \sum_{i,j}^{M,N} \bm{\gamma}_{i,j}\log(\bm{W}_{i,j}+\epsilon)  \\
&\qquad \qquad \mbox{s.t.}  \quad 0 \le \bm{W}_{i,j} \le 1 
\end{aligned}
\end{equation}
By taking the derivative of above \eqref{Wsubproblem} to zero, the closed-form solution of (\ref{Wsubproblem}) can be formulated as 
\begin{equation} \label{updateW}
\hat{\bm{W}}_{i,j}^{t+1} = \left\{ 
\begin{aligned}
&\frac{\sqrt{\bm{\gamma}_{i,j}} \sigma}{\left|\bm{Y}_{i,j}^{t}-\bm{X}_{i,j}^{t}\right|+\epsilon}  &  \sqrt{\bm{\gamma}_{i,j}} \sigma< \left|\bm{Y}_{i,j}^{t}-\bm{X}_{i,j}^{t}\right| \\
& 1      & \sqrt{\bm{\gamma}_{i,j}} \sigma > \left|\bm{Y}_{i,j}^{t}-\bm{X}_{i,j}^{t}\right| 
\end{aligned}
\right.
\end{equation}
where the superscript $t$ denotes at the $t$-th iteration. For simplicity, the superscript $t$ is omitted without confusion.

\subsubsection{Solving the $\mathbf{X}$ subproblem}
Fixing $\bm{W}$ and $\sigma$,  the clean image $\bm{X}$ can be estimated by 
\begin{equation}\label{Xsubproblem}
\hat{\bm{X}} = \arg \min_{\bm{X}} \frac{1}{2\sigma^{2}} \left\| \bm{W} \odot (\bm{Y}-\bm{X})  \right\| _{F}^{2}+\lambda \sum_{p=1}^{C} \left\| \bm{R}_{p}\bm{X}\right\|_{*} 
\end{equation}
To facilitate the optimization, we introduce an auxiliary variables  $\bm{U}$, \ie, $\bm{U} = \bm{X}$. Therefore, (\ref{Xsubproblem}) can be rewritten

\begin{equation}\label{Xsubproblem2}
\begin{aligned}
\hat{\bm{X}} = \arg \min_{\bm{X}} \frac{1}{2\sigma^{2}} \left\| \bm{W} \odot (\bm{Y}-\bm{X})  \right\| _{F}^{2}+\lambda \sum_{p=1}^{C} \left\| \bm{R}_{p} \bm{U}\right\|_{*}  \\
s.t. \quad \bm{U} = \bm{X}
\end{aligned}
\end{equation}
By enforcing the constraint with the Bregman iteration process \cite{goldstein2009split}, the minimization problem (\ref{Xsubproblem2}) becomes 
\begin{equation} \label{Xsubproblem3}
\begin{aligned}
(\hat{\bm{X}},\hat{\bm{U}}) = \arg &\min_{\bm{X},\bm{U}} \frac{1}{2\sigma^{2}} \left\| \bm{W} \odot (\bm{Y}-\bm{X})  \right\| _{F}^{2} \\
&+\lambda \sum_{p=1}^{C} \left\|  \bm{R}_{p}\bm{U}\right\|_{*} 
+   \frac{\beta}{2}\left\|   \bm{X} - \bm{U} - \bm{B} \right\|_{F}^{2}
\end{aligned}
\end{equation}
where $\bm{B}$ is a new auxiliary variabler introduced by splitting Bregman iterations \cite{goldstein2009split}, and $\beta$ is a regularization parameter. Hence the optimization problem (\ref{Xsubproblem3}) can be split into two subproblems: 

\begin{align}
&\hat{\bm{X}} = \arg \min_{\bm{X}} \frac{1}{2 \sigma^{2}} \left\|\bm{W} \odot (\bm{Y}-\bm{X}) \right\| _{F}^{2}+ \frac{\beta}{2}\left\|   \bm{X} - \bm{U} - \bm{B} \right\|_{F}^{2} \label{Xsubproblem4} \\
&\hat{\bm{U}}  =  \arg \min_{\bm{U}} \frac{\beta}{2}\left\|   \bm{X} - \bm{U} - \bm{B} \right\|_{F}^{2} +\lambda  \sum_{p=1}^{C} \left\| \bm{R}_{p}\bm{U}\right\|_{*} \label{Usubproblem}
\end{align}


Then the $\bm{X}$ subproblem in (\ref{Xsubproblem4}) can be easily solved in a closed-form solution, that is, 
\begin{equation} \label{updateX}
\begin{aligned}
\hat{\bm{X}} = \left [  \bm{W}\odot\bm{W}\odot \bm{Y}+ \beta \sigma^{2} (\bm{U}+\bm{B})\right ]   \oslash  \left(\bm{W}\odot\bm{W}+ \beta \sigma^{2}\right) 
\end{aligned}
\end{equation}
Where $\oslash$ denotes the element-wise division.

For solving the $\bm{U}$ subproblem in (\ref{Usubproblem}), instead of directly obtain $\bm{U}$, we adopt the strategy widely used in patch based method \cite{dabov2007image, mairal2009non}, that is, we first update groups of the similar patches $\bm{R}_{p}\bm{U}$, and then reconstruct the whole $\bm{U}$. Recall that we assume each group of similar patches are independent in (\ref{lowrank}). Let $\bm{L} = \bm{X}-\bm{B}$  and we have the following conclusion:

\begin{algorithm}[t] 
	\caption{Proposed RWE method via solving (\ref{Propose2})}
	\begin{itemize} 
		\item \textbf{Input}: Noisy image $\bm{Y}$
		\item Initialize $\bm{X}^{0}$ via adaptive median filter. 
		\item Set the parameters $\beta, \theta$, the maximum iteration $I_{max}$;
		\item Construct the similar patches collected operator $\bm{R_{j}}$ by kNN (k-NearestNeighbor) algorithm.
		\item Outer loop $t=1,...,I_{max}$ 
		\begin{itemize}
			\item Update $\sigma^{t}$   based on Eq.(\ref{updatesigma}).
			\item Update parameter $\bm{\gamma}_{i,j}^{t}$ based on Eq.(\ref{updategamma}).
			\item Update $\bm{W}^{t}$  based on Eq.(\ref{updateW}).
			\\
			Inner loop $l = 1, ... , C $ for solving $\bm{X}$ subproblem
			\begin{itemize}
				\item[-] Update $\bm{X}$ based on Eq.(\ref{updateX}).
				\item[-] Update $\bm{U}$ based on Eq.(\ref{SVD}) and Eq.(\ref{updateU}).
				\item[-] Update $\bm{B}$ based on Eq.(\ref{updateB}).
			\end{itemize}
			end loop 
			\item If mod (iter,4) = 0, update the operator $\bm{R}_{j}$;
		\end{itemize}
		end loop 
		\item \textbf{Output}: Estimated image $\bm{X}$
	\end{itemize} 
	\label{algorithm1} 
\end{algorithm}

\begin{table}[h] 
	\centering
	\caption{The values of $E1$ and $E2$ at different iterations in the optimization process.}
	\label{Tab:equi}
	\begin{tabular}{ccccc}
		\hline \hline
		Iterations & 1    & 3    & 5    & 7    \\ \hline 
		$E1$         & 4.09$\times 10^{-4}$ & 3.93$\times 10^{-4}$ & 3.57$\times 10^{-4}$ & 3.04$\times 10^{-4}$ \\ \hline
		$E2$         & 3.78$\times 10^{-4}$ & 3.67$\times 10^{-4}$ & 3.31$\times 10^{-4}$ & 2.92$\times 10^{-4}$ \\ \hline \hline
	\end{tabular}
\end{table}

\begin{figure}[h]
	\setcounter{subfigure}{0} 
	\centering
	\subfloat{
		\begin{minipage}{0.245\textwidth}
			\centering
			\setlength{\abovecaptionskip}{-0.1cm}
			\setlength{\belowcaptionskip}{-0.1cm}
			\includegraphics[width=1\columnwidth]{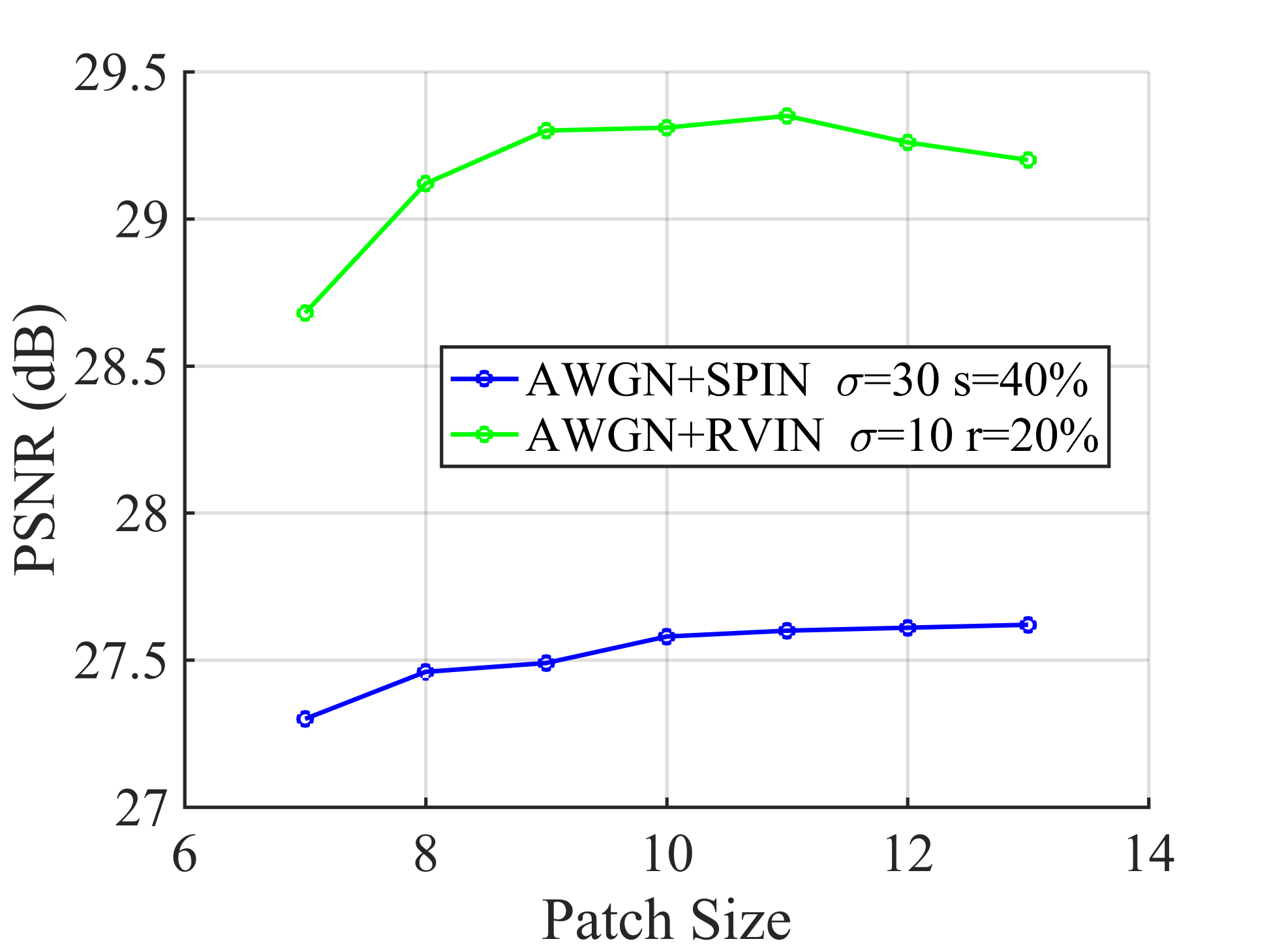}
			\label{parameters:a}
			\caption*{(a)}
		\end{minipage}
	}
	\subfloat{
		\begin{minipage}{0.245\textwidth}
			\centering
			\setlength{\abovecaptionskip}{-0.1cm}
			\setlength{\belowcaptionskip}{-0.1cm}
			\includegraphics[width=1\columnwidth]{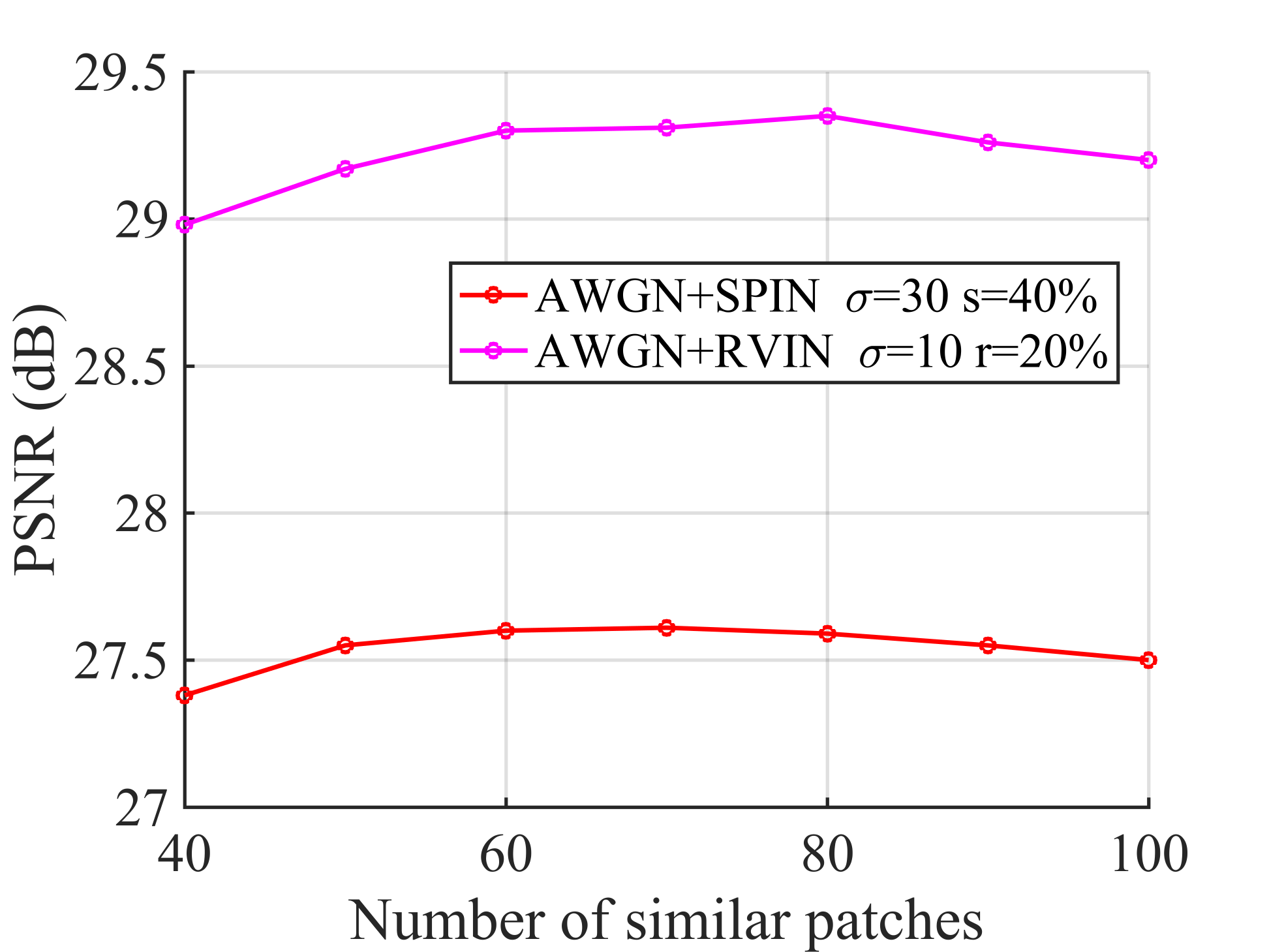}
			\label{parameters:b}
			\caption*{(b)}
		\end{minipage}
	}  	
	\caption{The PSNR curves with  (a) different patch size $\sqrt{m}$; (b) different number of similar patches $K$}
	\label{parametersetting}
\end{figure}

\begin{figure}[h]
	\setcounter{subfigure}{0} 
	\centering
	\setlength{\abovecaptionskip}{-0.05cm}
	\setlength{\belowcaptionskip}{0.1cm}
	\subfloat{
		\begin{minipage}{0.245\textwidth}
			\centering
			\includegraphics[width=1\columnwidth]{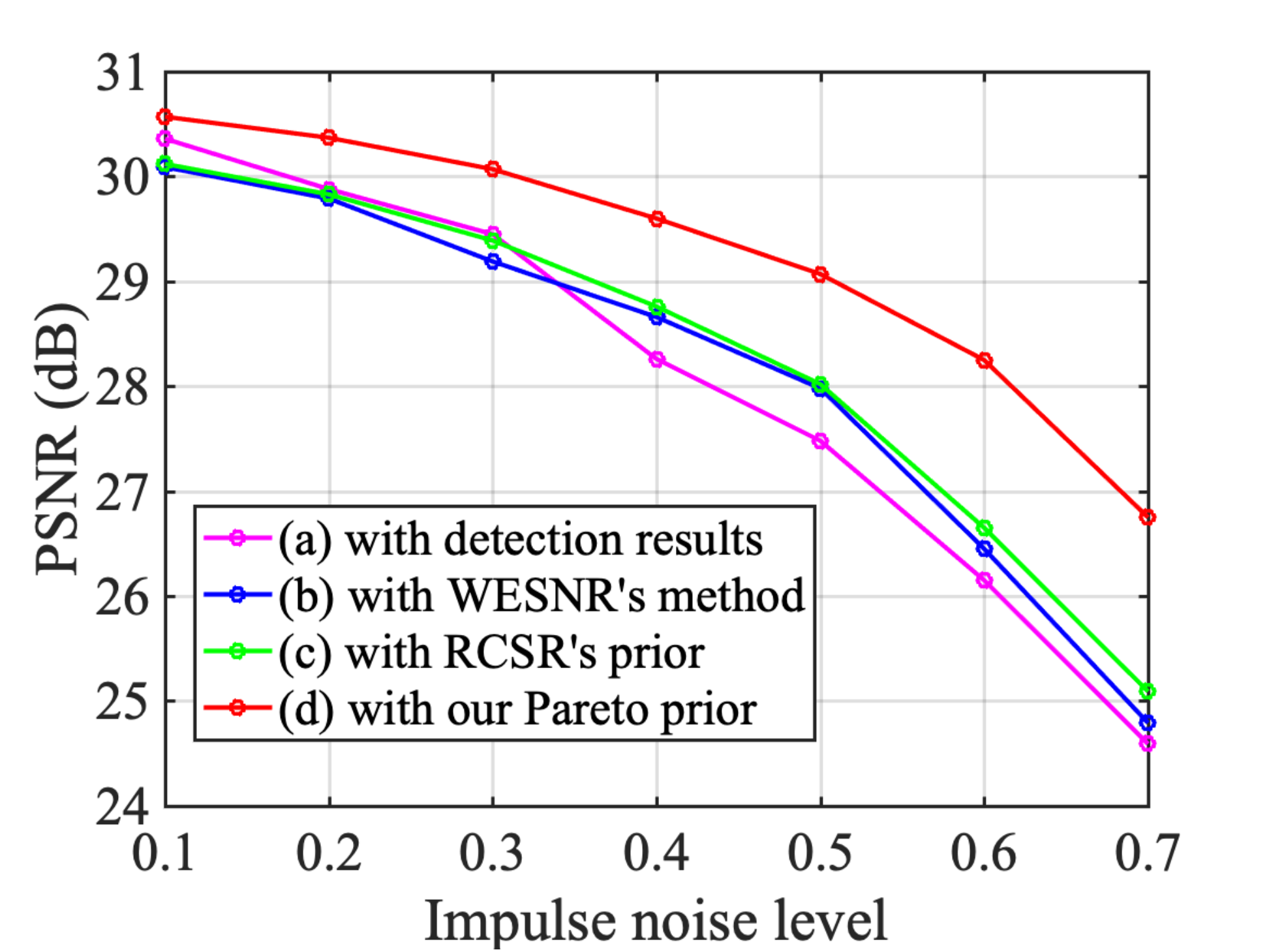}
			\label{effectiveness:lena}
			\caption*{(a)}
		\end{minipage}
	}
	\subfloat{
		\begin{minipage}{0.245\textwidth}
			\centering
			\setlength{\abovecaptionskip}{-0.05cm}
			\setlength{\belowcaptionskip}{0.1cm}
			\includegraphics[width=1\columnwidth]{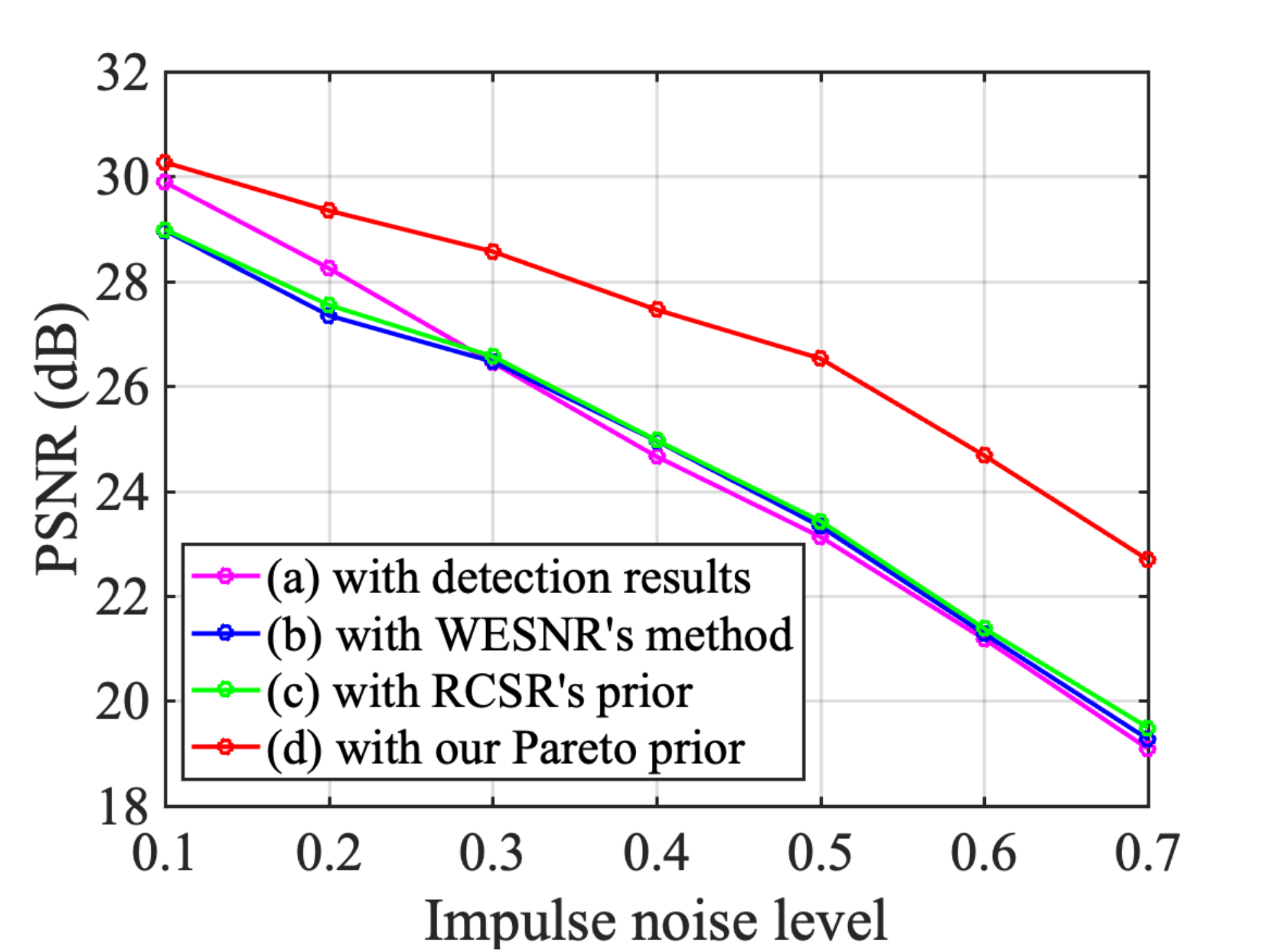}
			\label{effectiveness:barbara}
			\caption*{(b)}
		\end{minipage}
	}  	
	\caption{The PSNR curves with various mixed noise levels on test images:  (a) \textit{Lena} (AWGN+SPIN: $\sigma =30$, $s = [0.1,0.7]$) and (b) \textit{Barbara} (AWGN+RVIN: $\sigma = 10$, $r=[0.1,0.7]$). }
	\label{effectiveness}
\end{figure}

\begin{figure*}[ht]
	\setcounter{subfigure}{0} 
	\centering
	\subfloat{
		\begin{minipage}{0.16\textwidth}
			\centering
				\setlength{\abovecaptionskip}{-0.1cm}
			\setlength{\belowcaptionskip}{-0.1cm}
			\includegraphics[width=1\columnwidth]{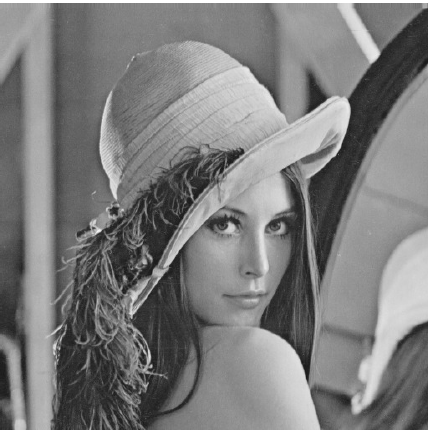}
			\label{lena:original}
\caption*{(a) Orignial image\\\hspace*{.4cm}~~}
		\end{minipage}
	}  	
	\subfloat{
		\begin{minipage}{0.16\textwidth}
			\centering
				\setlength{\abovecaptionskip}{-0.2cm}
			\setlength{\belowcaptionskip}{0cm}
			\includegraphics[width=1\columnwidth]{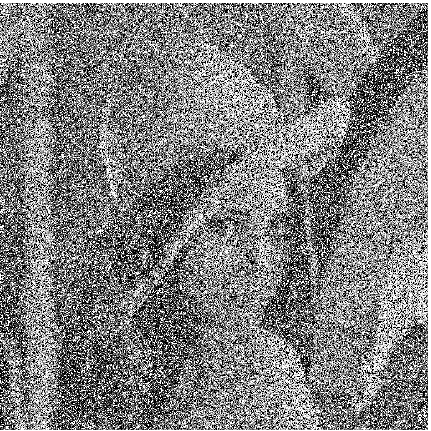}
			\label{lena:noisy}
			\caption*{\centerline{(b) Noisy image} \\\hspace*{.4cm} (PSNR(dB)/SSIM)}
		\end{minipage}
	}
	\subfloat{
		\begin{minipage}{0.16\textwidth}
			\centering
				\setlength{\abovecaptionskip}{-0.2cm}
			\setlength{\belowcaptionskip}{0cm}
			\includegraphics[width=1\columnwidth]{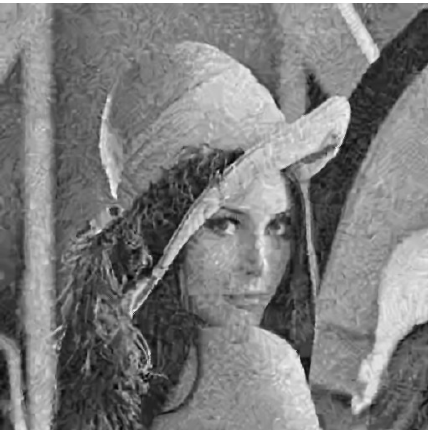}
			\label{lena:BM3D}
			\caption*{\centerline{(c) MBM3D \cite{dabov2007image}}\\\hspace*{.6cm}(26.35/65.57\%)}
		\end{minipage}
	}  	
	\subfloat{
		\begin{minipage}{0.16\textwidth}
			\centering
				\setlength{\abovecaptionskip}{-0.2cm}
			\setlength{\belowcaptionskip}{0cm}
			\includegraphics[width=1\columnwidth]{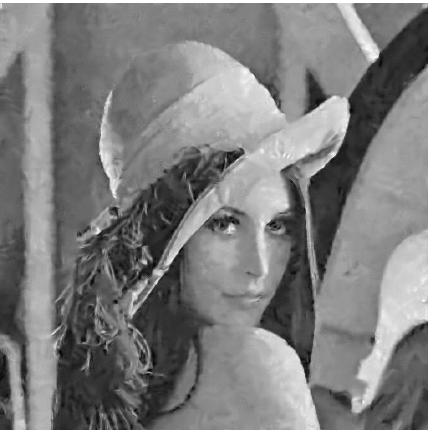}
			\label{lena:WESNR}
			\caption*{\centerline{(d) WESNR \cite{jiang2014mixed}}\\\hspace*{.6cm}(27.79/76.53\%)}
		\end{minipage}
	}  	 
	\subfloat{
		\begin{minipage}{0.16\textwidth}
			\centering
				\setlength{\abovecaptionskip}{-0.2cm}
			\setlength{\belowcaptionskip}{0cm}
			\includegraphics[width=1\columnwidth]{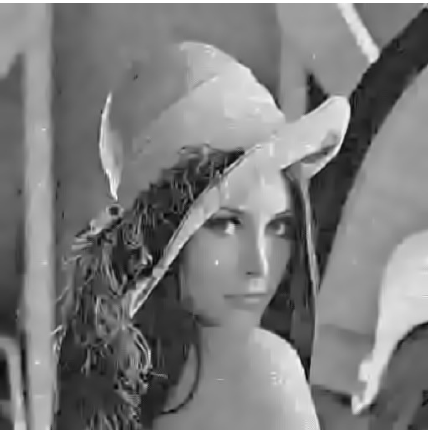}
			\label{lena:WJSR}
			\caption*{ \centerline{(e) WJSR \cite{liu2018mixed}}\\\hspace*{.7cm}(27.94/78.74\%)}
		\end{minipage}
	}  	
	\subfloat{
		\begin{minipage}{0.16\textwidth}
			\centering
				\setlength{\abovecaptionskip}{-0.2cm}
			\setlength{\belowcaptionskip}{0cm}
			\includegraphics[width=1\columnwidth]{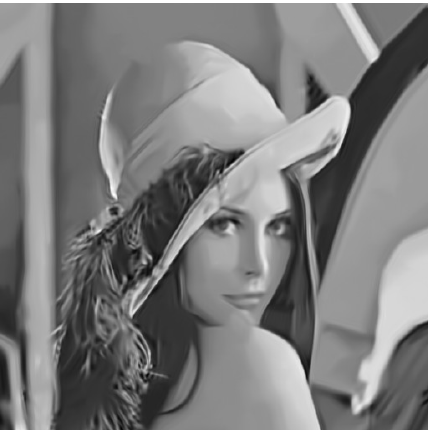}
			\label{lena:Proposed}
			\caption*{\centerline{(f) Our RWE}\\\hspace*{.6cm}(\textbf{29.07}/ \textbf{81.77}\%)}
				\end{minipage}
	}  	
	\caption{Denoising on test image \textit{Lena} corrupted with AWGN+SPIN ($\sigma = 30$, $s = 50\%$). }
	\label{Exp:lena}
\end{figure*}

\begin{figure*}[ht] \vspace*{-.6cm}
	\setcounter{subfigure}{0} 
	\centering
	
	\subfloat{
		\begin{minipage}{0.16\textwidth}
			\centering
			\setlength{\abovecaptionskip}{-0.2cm}
			\setlength{\belowcaptionskip}{0cm}
			\includegraphics[width=1\columnwidth]{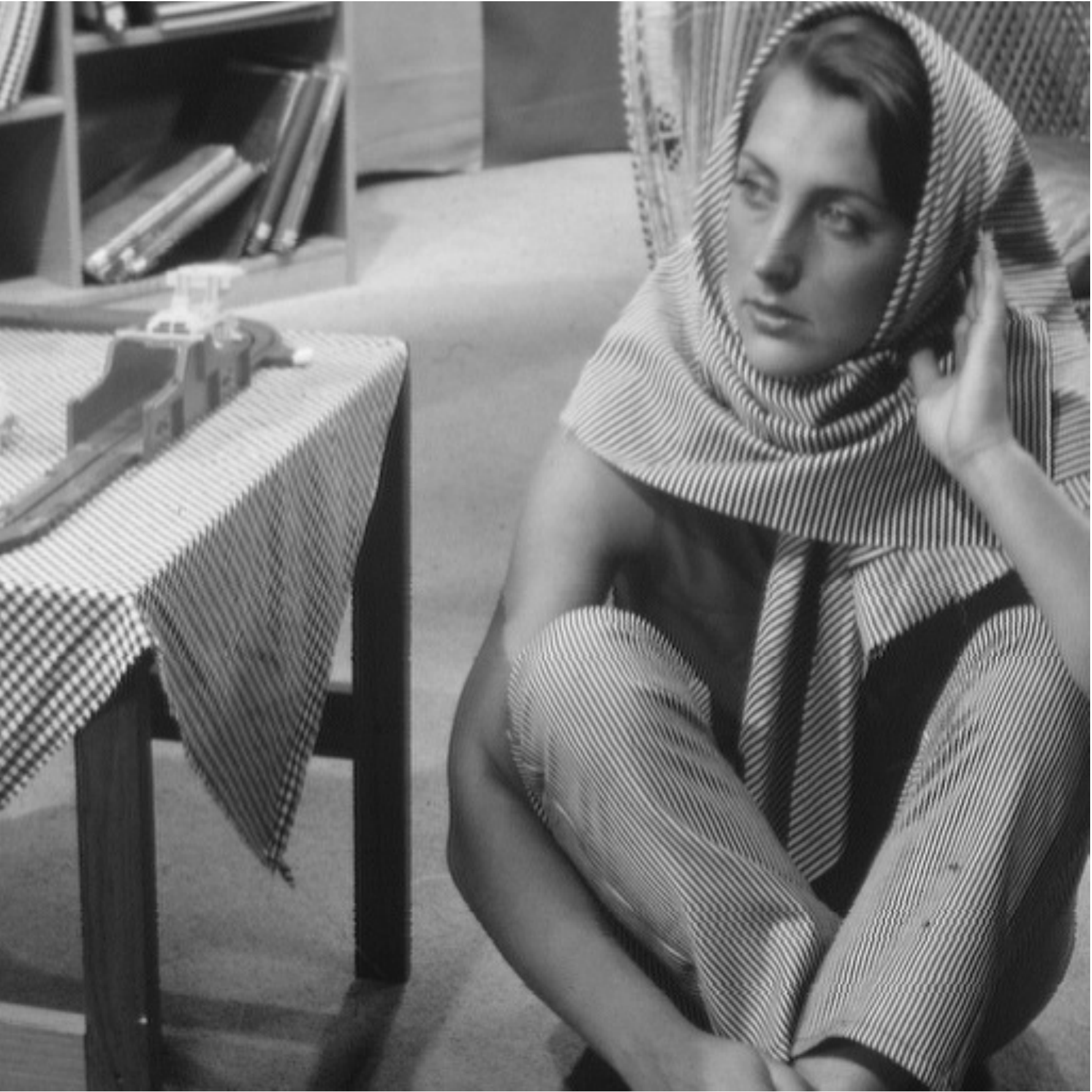}
			\label{barbara:original}
			\caption*{(a) Oringal image \\\hspace*{.6cm}}
		\end{minipage}
	}  		
	\subfloat{
		\begin{minipage}{0.16\textwidth}
			\centering
			\setlength{\abovecaptionskip}{-0.2cm}
			\setlength{\belowcaptionskip}{0cm}
			\includegraphics[width=1\columnwidth]{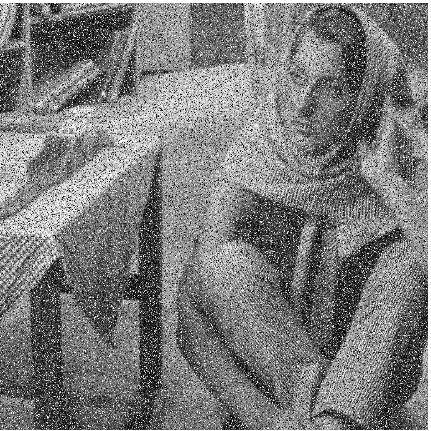}
			\label{barbara:noisy}
			\caption*{\centerline{(b) Noisy image} \\\hspace*{.4cm} (PSNR(dB)/SSIM)}
		\end{minipage}
	}
	\subfloat{
		\begin{minipage}{0.16\textwidth}
			\centering
			\setlength{\abovecaptionskip}{-0.2cm}
			\setlength{\belowcaptionskip}{0cm}
			\includegraphics[width=1\columnwidth]{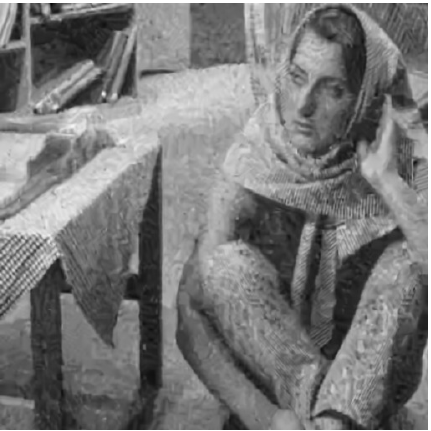}
			\label{barbara:BM3D}
			\caption*{\centerline{(c) MBM3D \cite{dabov2007image}}\\\hspace*{.6cm}(22.94/61.35\%)}
		\end{minipage}
	}  	
	\subfloat{
		\begin{minipage}{0.16\textwidth}
			\centering
			\setlength{\abovecaptionskip}{-0.2cm}
			\setlength{\belowcaptionskip}{0cm}
			\includegraphics[width=1\columnwidth]{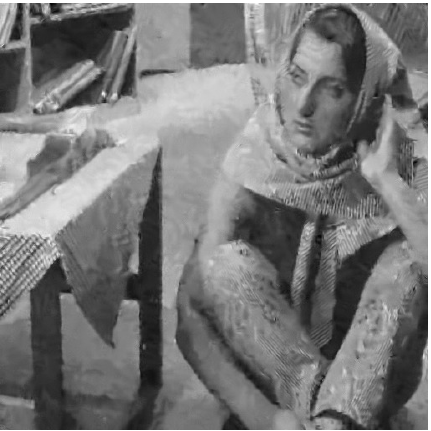}
			\label{barbara:WESNR}
			\caption*{\centerline{(d) WESNR \cite{jiang2014mixed}}\\\hspace*{.6cm}(23.29/65.00\%)}
		\end{minipage}
	}  	 
	\subfloat{
		\begin{minipage}{0.16\textwidth}
			\centering
			\setlength{\abovecaptionskip}{-0.2cm}
			\setlength{\belowcaptionskip}{0cm}
			\includegraphics[width=1\columnwidth]{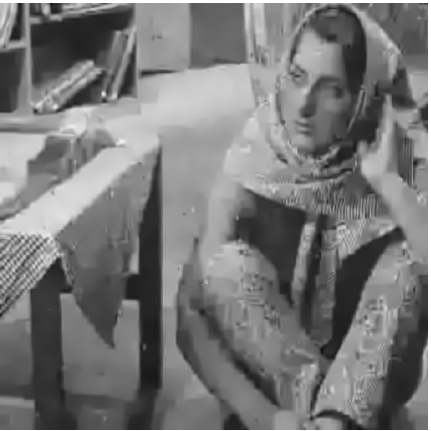}
			\label{barbara:WJSR}
			\caption*{ \centerline{(e) WJSR \cite{liu2018mixed}}\\\hspace*{.7cm}(22.98/63.20\%)}
		\end{minipage}
	}  	
	\subfloat{
		\begin{minipage}{0.16\textwidth}
			\centering
			\setlength{\abovecaptionskip}{-0.2cm}
			\setlength{\belowcaptionskip}{0cm}
			\includegraphics[width=1\columnwidth]{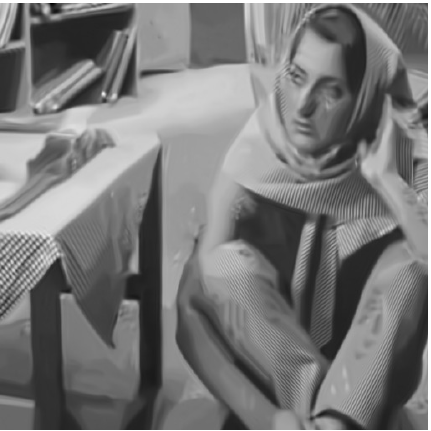}
			\label{barbara:Proposed}
			\caption*{\centerline{(f) Our RWE}\\\hspace*{.6cm}(\textbf{24.43}/\textbf{71.00}\%)}
		\end{minipage}
	}  	
    \caption{Denoising on test image \textit{Barbara} corrupted with AWGN+RVIN ($\sigma = 30$, $r = 30\%$).}
	\label{Exp:barbara}
\end{figure*}

\begin{thm}\cite{zhang2014group} \textbf{:}
Let $\bm{L},\bm{U} \in \mathbb{R}^{M \times N}$,  $\bm{R}_{p}\bm{L},\bm{R}_{p}\bm{U}\in \mathbb{R}^{m \times K}$, where $\bm{R}_{p}$ denotes the operator extracting the similar patches, $G = M N$,  $D = mKT$, $\bm{Z} = \bm{L} - \bm{U}$ and denote each element of $\bm{Z}$ by $\bm{Z}_{i,j}$, $i = 1,2,...,M$ and $j=1,2,...,N$. Assume that $\bm{Z}_{i,j}$ is independent and follows a distribution with zero-mean and variance $\sigma_{w}^{2}$. For any $\epsilon >0$, we have the following result,
\begin{equation} \label{theoremlaw}
\begin{aligned}
&\lim _{ \substack{G \rightarrow \infty \\  D \rightarrow \infty} } P\left\{\left|\frac{1}{G}\|\bm{L}-\bm{U}\|_{2}^{2}
-\frac{1}{D} \sum_{p=1}^{C}\left\|\bm{R}_{p}\bm{L}-\bm{R}_{p}\bm{U}\right\|_{F}^{2}\right|<\varepsilon\right\}\\
&\qquad \qquad  \qquad  \qquad   \qquad  \qquad  \qquad  \qquad  \qquad \qquad =1
\end{aligned}
\end{equation}
\end{thm}

According to Theorem 1,  we have the following equation with very large probability (limited to 1) 
\begin{equation} \label{equivalence}
\frac{1}{MN}\|\bm{X}-\bm{B}-\bm{U}\|_{2}^{2}= \frac{1}{mKT} \sum_{p=1}^{C}\left\|\bm{R}_{p}\bm{X}-\bm{R}_{p}\bm{B}-\bm{R}_{p}\bm{U}\right\|_{F}^{2}
\end{equation}
Furthermore,  we can verify (\ref{equivalence}) at each iteration in optimization process. We denote the left hand of Eq.(\ref{equivalence}) by $E1$ and the right hand of Eq.(\ref{equivalence}) by $E2$. In Table~\ref{Tab:equi}. , we present $E1$ and $E2$ at different iterations on  test image \textit{Barbara} corrupted with AWGN+SPIN ($\sigma$ = 10, s = 20\%). It can be seen clearly that $E1$ is very close to $E2$, which sufficiently illustrates the validity of Eq.(\ref{equivalence}).

By substituting Eq. (\ref{equivalence}) into (\ref{Usubproblem}),  we have 
\begin{equation} \label{Usubproblem2}
\min_{\bm{R}_{p}\bm{U}} \frac{\alpha}{2} \sum_{p=1}^{C}\left\|\bm{R}_{p}\bm{X}-\bm{R}_{p}\bm{B}-\bm{R}_{p}\bm{U}\right\|_{F}^{2}+\lambda  \sum_{p=1}^{C} \left\| \bm{R}_{p}\bm{U}\right\|_{*}
\end{equation}
where $\alpha = \frac{\beta MN}{mK}$. The above \eqref{Usubproblem2} is a standard nuclear norm minimization and can be solved effectively by singular value thresholding \cite{cai2010singular}:
\begin{equation} \label{SVD}
\bm{R}_{p}\hat{\bm{U}} = \bm{H}S(\bm{\Sigma},\theta)\bm{V}^{T}
\end{equation}
where $S(v, \theta)$ is the element-wise soft-thresholding operator with threshold $\theta= \frac{\lambda}{\alpha}$, i.e., $ S(v, \theta) = sgn(v) \times \max(|v|- \theta, 0)$ and $(\bm{H},\bm{\Sigma},\bm{V})= \mbox{SVD} (\bm{R}_{p}\bm{X}-\bm{R}_{p}\bm{B})$, SVD denotes the operator of singular value decomposition. Subsequently, the whole $\bm{U}$ can be obtained by  
\begin{equation} \label{updateU}
\bm{U} = \frac{1}{a} \sum_{p=1}^{C} \bm{R}_{p}^{T}\bm{R}_{p}\bm{U}
\end{equation}
where $a$ is the normalization factor. The $\bm{B}$ can be updated by 
\begin{equation}\label{updateB}
\bm{B}^{l+1} = \bm{B}^{l} + (\bm{U} - \bm{X})
\end{equation}
Where $\bm{B}^{l}$ denotes $\bm{B}$ obtained in the $l$-th iteration.

\subsubsection{Solving the $\sigma$ subproblem}
For fixing $\bm{X}$ and $\bm{W}$, the variance of AWGN can be estimated by solving the following problem:
\begin{equation}\label{sigmasubproblem}
\hat{\sigma} = \arg \min_{\sigma} \frac{1}{2\sigma^{2}} \left\| \bm{W} \odot (\bm{Y}-\bm{X})  \right\| _{F}^{2}+  MN\log\sigma 
\end{equation}
The closed-form solution to $\sigma$ can be directly obtained by setting the derivative of above  \eqref{sigmasubproblem} to zero.
Hence $\sigma$ can be updated by  
\begin{equation}\label{updatesigma}
\hat{\sigma} = \sqrt{ \frac{\left\|\bm{W} \odot (\bm{Y}-\bm{X})\right\|_{F}^{2} }{MN} }
\end{equation}

Up to now, the efficient solution for each minimization subproblem has been acquired. A detailed descriptions of our proposed algorithm for mixed noise removal are provided in Algorithm~\ref{algorithm1}.

\begin{table*}[h]
	\fontsize{6.5pt}{\baselineskip}\selectfont
	\centering
	\caption{\footnotesize The estimated means and variances of weighted residuals comparison with different methods.}
	\label{tab:robust}
\begin{tabular}{ccccccccccccc}
		\hline \hline
		& \multicolumn{4}{c}{AWGN+SPIN}                                                       & \multicolumn{4}{c}{AWGN+RVIN}                                                       & \multicolumn{4}{c}{AWGN+SPIN+RVIN}                                                                     \\
		& \multicolumn{2}{c}{$\sigma$ = 10, $s$ = 30\%} & \multicolumn{2}{c}{$\sigma$  = 30, $s$ = 70\%} & \multicolumn{2}{c}{$\sigma$ = 10, $r$ = 30\%} & \multicolumn{2}{c}{$\sigma$  = 50, $r$ = 50\%} & \multicolumn{2}{c}{$\sigma$  = 10, $s$ = 30\%, $r$ = 30\%} & \multicolumn{2}{c}{$\sigma$  =30, $s$ = 50\%, r = 50\%} \\ \cline{2-13} 
		& Mean               & Variance            & Mean               & Variance            & Mean               & Variance            & Mean               & Variance            & Mean                    & Variance                 & Mean                   & Variance                 \\
		WESNR \cite{jiang2014mixed}       & 0.0104             & 7.69                & 0.6116             & 28.33               & 0.0432             & 12.96               & 0.6692             & 56.46               & 0.2026                  & 15.54                    & 0.4473                 & 34.1128                  \\
		WJSR \cite{liu2017weighted}        & 0.0064             & 7.21                & 0.2154             & 27.23               & \textbf{0.0069}             & 8.34                & 0.1563             & 21.52               & -                       & -                        & -                      & -                        \\
		RCSR \cite{liu2018mixed}        & 0.0063             & 5.71                & 0.5882             & 22.82               & 0.0319             & 8.75                & 0.6651             & 25.05               & 0.2140                  & 6.47                     & 0.5865                 & 18.97                    \\
\hline \hline
	\end{tabular}
	
\end{table*}

\begin{table*}[h]
	\tiny
	\centering
	\caption{The PSNR(dB)/SSIM(\%) results of mixed noise removal (AWGN+SPIN) on 12 standard test images.}
	\label{Exp:SPIN}
	\begin{tabular}{cccccc|cccc|cccc}
		\hline \hline 
		\multicolumn{2}{c}{}                                 & \multicolumn{4}{c|}{$\sigma = 10$}                                                                            & \multicolumn{4}{c|}{$\sigma=30$}                                                                              & \multicolumn{4}{c}{$\sigma=50$}                                                                              \\ \hline
		Image                                           &   s   & \multicolumn{1}{c}{MBM3D \cite{dabov2007image}} & \multicolumn{1}{c}{WESNR \cite{jiang2014mixed}} & \multicolumn{1}{c}{WJSR \cite{liu2017weighted}} & \multicolumn{1}{c|}{RWE(ours)} & \multicolumn{1}{c}{MBM3D \cite{dabov2007image}} & \multicolumn{1}{c}{WESNR \cite{jiang2014mixed}} & \multicolumn{1}{c}{WJSR \cite{liu2017weighted}} & \multicolumn{1}{c|}{RWE (ours)} & \multicolumn{1}{c}{MBM3D \cite{dabov2007image}} & \multicolumn{1}{c}{WESNR \cite{jiang2014mixed}} & \multicolumn{1}{c}{WJSR \cite{liu2017weighted}} & \multicolumn{1}{c}{RWE (ours)} \\ \hline
		\multirow{5}{*}{Airplane}                       & 20\% & 32.36/91.41               & 33.06/90.06                & 33.50/91.47               & \textbf{33.97/92.07}     & 28.44/80.40               & 28.36/83.62                & 28.93/84.27               & \textbf{29.36/85.67}     & 26.00/73.28               & 18.82/49.78                & 26.52/78.81               & \textbf{26.96/81.29}     \\ 
		& 30\% & 30.81/90.22               & 32.41/89.37                & 33.07/91.17               & \textbf{33.76/91.78}     & 27.53/76.56               & 27.75/81.71                & 28.22/83.00               & \textbf{28.85/84.95}     & 25.29/69.09               & 17.02/43.12                & 25.60/77.33               & \textbf{26.42/80.46}     \\
\hline \hline
	\end{tabular}
\end{table*}

\subsection{Adaptive Updating Shape Parameter} \label{Subsec:gamma}

For  $\bm{\gamma}_{i,j}$ in Eq.(\ref{Wsubproblem}), we avoid to set it empirically. Inspired by robust statistics \cite{maronna2006robust},  we propose to update $\bm{\gamma}_{i,j}$ adaptively using the residual map, which is defined as follows: 

\begin{equation}
\bm{S}^{t} = \bm{Y} - \bm{X}^{t-1}
\end{equation}
where $\bm{S}^{t}$ is the residual map at the $t$-th iteration and  $\bm{X}^{t-1}$ is the estimated image at the $(t-1)$-th iteration. Then we can calculate the parameter $\gamma_{i,j}$ 
\begin{equation}\label{updategamma}
\bm{\gamma}_{i,j}^{t} = e^{-\text{MAD}(\bm{S}^{t}_{i,j})}
\end{equation}
where $\text{MAD}(\cdot)$ stands for the operator of \textit{median absolute deviation about the median} (MAD), which is a robust alternative to the standard deviation. We calculate MAD as follows 
\begin{equation}
\text{MAD}(z) = Med(|z - Med(z)|)
\end{equation}
Where $Med$ stands for the operator of median value. In this way, $\bm{\gamma}_{i,j}$ can be updated in an adaptive manner. When mixed noise level is high, we have a larger MAD value, leading to a small $\bm{\gamma}_{i,j}$, and vice versa. 


\subsection{Convergence Analysis} \label{Sec:convergence}


In total, there are two parameters in optimization process need to be tuned, \ie, $\beta$, $\theta$. In particular, $\beta$ is a regularization parameter introduced by splitting Bregman itertation \cite{goldstein2009split} in Eq.(\ref{Xsubproblem3}). We choose $\beta = 5$ because of the good convergence. For the threshold parameter $\theta$ in Eq.(\ref{SVD}), we follow the suggested setting of \cite{dong2013nonlocal}, \ie, $\theta= 2 \sqrt{2}\sigma^{2}$. 

%

For $\bm{X}$ subproblem, the convergence can be guaranteed in theory by the splitting Bregman iterations \cite{goldstein2009split}. 
Note that $\bm{W}$ and $\bm{X}$ subproblems are convex, but $\sigma$ subproblem in (\ref{sigmasubproblem}) is non-convex. Thus strictly mathematical proof of the convergence of our overall algorithm is difficult. However, the convergence of our algorithm can be analyzed briefly as follows: Denote by $F(\bm{X},\bm{W},\sigma)$ the objective function in (\ref{Propose2}). In the $(i+1)$-th iteration, $\hat{\bm{W}}^{i+1}$ is first optimizied by solving the convex problem (\ref{Wsubproblem}). Then $\hat{\bm{X}}^{i+1}$ can achieve the optimal solution by solving (\ref{Xsubproblem3}). Hence we have $F(\bm{X}^{i+1},\bm{W}^{i+1},\sigma^{i}) \le F(\bm{X}^{i},\bm{W}^{i},\sigma^{i})$. When the local optimal point $\sigma$ achieve by solving (\ref{sigmasubproblem}), we have $F(\bm{X}^{i+1},\bm{W}^{i+1},\sigma^{i+1}) \le F(\bm{X}^{i+1},\bm{W}^{i+1},\sigma^{i})$. Consequently, the following inequalities hold,
\begin{equation}
F(\bm{X}^{max},\bm{W}^{max},\sigma^{max}) \le ... \le F(\bm{X}^{1},\bm{W}^{1},\sigma^{1})
\end{equation}
The inequalities above illustrate that our algorithm will be convergent to a tolerance value after some appropriate iterations if a proper initialization is given. This is consistent with our empirical observation, as presented in Section~\ref{Exp:time},


\begin{figure*}[h] \vspace*{-.5cm}
	\setlength{\abovecaptionskip}{0cm}
	\setlength{\belowcaptionskip}{-0.2cm}
	\setcounter{subfigure}{0} 
	\centering
	\subfloat{
		\begin{minipage}{0.19\textwidth}
			\centering
			\setlength{\abovecaptionskip}{-0.2cm}
			\setlength{\belowcaptionskip}{0cm}
			\includegraphics[width=1\columnwidth]{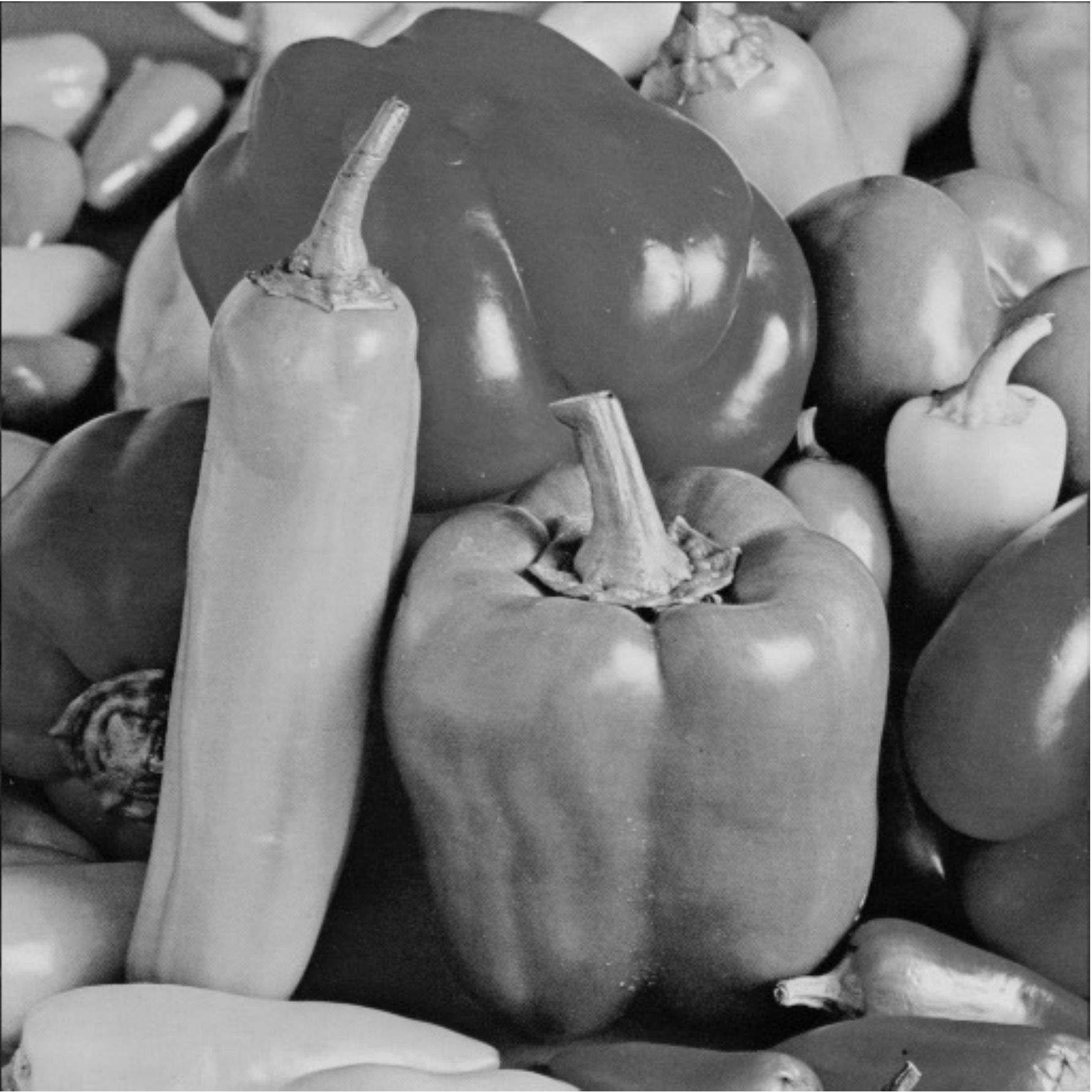}
			\label{peppers:original}
			\caption*{(a) Original image \\\hspace*{.3cm}}
		\end{minipage}
	}  	
	\subfloat{
		\begin{minipage}{0.19\textwidth}
			\centering
			\setlength{\abovecaptionskip}{-0.2cm}
			\setlength{\belowcaptionskip}{0cm}
			\includegraphics[width=1\columnwidth]{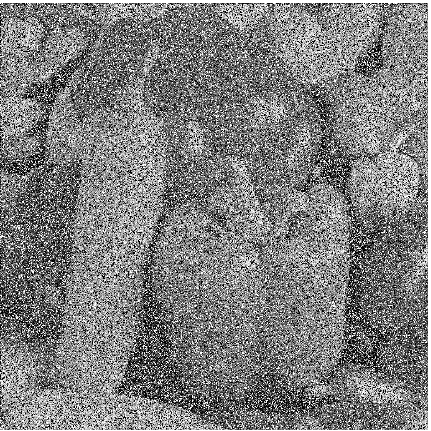}
			\label{peppers:noisy}
		\caption*{\centerline{(b) Noisy image} \\\hspace*{.6cm} (PSNR(dB)/SSIM)}
		\end{minipage}
	}
	\subfloat{
		\begin{minipage}{0.19\textwidth}
			\centering
			\setlength{\abovecaptionskip}{-0.2cm}
			\setlength{\belowcaptionskip}{0cm}
			\includegraphics[width=1\columnwidth]{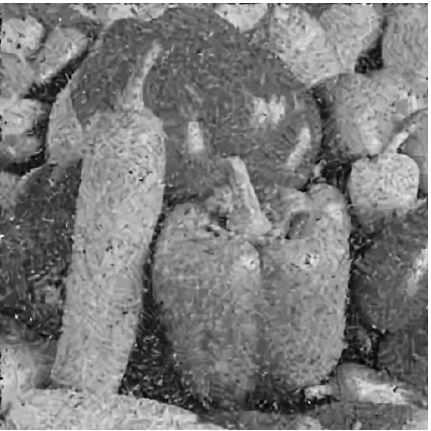}
			\label{peppers:BM3D}
\caption*{\centerline{(c) MBM3D \cite{dabov2007image}}\\\hspace*{.8cm}(21.00/47.06\%)}
		\end{minipage}
	}  	
	\subfloat{
		\begin{minipage}{0.19\textwidth}
			\centering
			\setlength{\abovecaptionskip}{-0.2cm}
			\setlength{\belowcaptionskip}{0cm}
			\includegraphics[width=1\columnwidth]{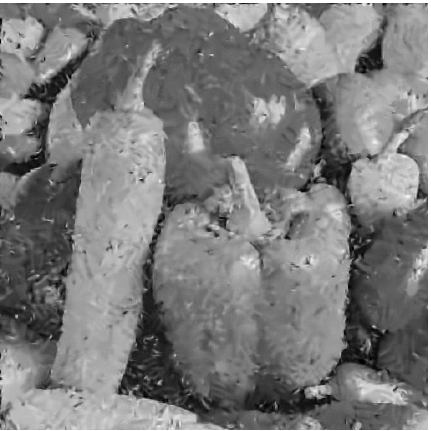}
			\label{peppers:WESNR}
			\caption*{\centerline{(d) WESNR \cite{jiang2014mixed}}\\\hspace*{.7cm}(22.29/59.37\%)}
		\end{minipage}
	}  	 
	\subfloat{
		\begin{minipage}{0.19\textwidth}
			\centering
			\setlength{\abovecaptionskip}{-0.2cm}
			\setlength{\belowcaptionskip}{0cm}
			\includegraphics[width=1\columnwidth]{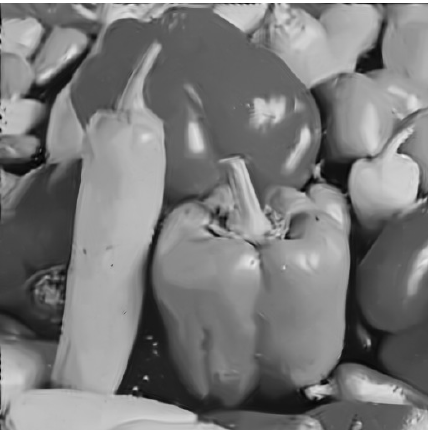}
			\label{peppers:Proposed}
			\caption*{\centerline{(f) Our RWE}\\\hspace*{.9cm}(\textbf{24.05}/\textbf{72.38}\%)}
		\end{minipage}
	}  	
		\caption{Denoising on test image \textit{Peppers} corrupted with AWGN+SPIN+RVIN ($\sigma = 30$, $s = 30\%$, $r= 40\%$). }
	\label{Exp:peppers}
\end{figure*}

\begin{table*}[h]
	\tiny
	\centering
	\caption{The PSNR(dB)/SSIM(\%) results of mixed noise removal (AWGN+RVIN) on 12 standard test images.}
	\label{Exp:RVIN}
\begin{tabular}{cccccc|cccc|cccc}
		\hline \hline
		\multicolumn{2}{c}{}                                                   & \multicolumn{4}{c|}{$\sigma = 10$}                                                                            & \multicolumn{4}{c|}{$\sigma=30$}                                                                              & \multicolumn{4}{c}{$\sigma=50$}                                                                              \\ \hline
		Image                                           & \multicolumn{1}{c}{r} & \multicolumn{1}{c}{MBM3D \cite{dabov2007image}} & \multicolumn{1}{c}{WESNR \cite{jiang2014mixed}} & \multicolumn{1}{c}{WJSR \cite{liu2017weighted}} & \multicolumn{1}{c|}{RWE (ours)} & \multicolumn{1}{c}{MBM3D \cite{dabov2007image}} & \multicolumn{1}{c}{WESNR \cite{jiang2014mixed}} & \multicolumn{1}{c}{WJSR \cite{liu2017weighted}} & \multicolumn{1}{c|}{RWE (ours)} & \multicolumn{1}{c}{MBM3D \cite{dabov2007image}} & \multicolumn{1}{c}{WESNR \cite{jiang2014mixed}} & \multicolumn{1}{c}{WJSR \cite{liu2017weighted}} & \multicolumn{1}{c}{RWE (ours)} \\ \hline
		\multirow{5}{*}{Airplane}                       & 20\%                   & 30.73/90.20               & 31.12/85.75                & 31.18/86.70               & \textbf{31.72/90.87}     & 27.15/77.84               & 27.38/81.51                & 27.30/82.15               & \textbf{27.95/83.92}     & 23.80/65.84               & 23.10/59.69                & 23.18/73.33               & \textbf{24.29/76.64}     \\ 
		& 30\%                   & 28.36/85.24               & 29.32/81.70                & 29.28/80.86               & \textbf{30.19/89.05}     & 25.69/72.92               & 26.23/79.08                & 26.20/79.98               & \textbf{26.91/82.21}     & 22.38/62.12               & 21.91/55.80                & 22.23/71.99               & \textbf{23.02/75.16}     \\
\hline \hline
	\end{tabular}
\end{table*}

\begin{table}[ht]
	\footnotesize
	\centering
	\setlength{\abovecaptionskip}{0.cm}
	\setlength{\belowcaptionskip}{0.cm}
	\caption{The average PSNR(dB)/SSIM(\%) of mixed noise (AWGN+SPIN+RVIN) removal on 12 standard test images.}
	\label{Exp:SPRV}
	\begin{tabular}{cccc}
		\hline \hline
		Mixed Noise                        & MBM3D \cite{dabov2007image}        & WESNR \cite{jiang2014mixed}       & RWE (ours)                  \\ \hline
		$\sigma = 10$ $s=0.2$ $r=0.3$ & 24.55/72.04 & 26.13/75.43 & \textbf{27.01/79.80} \\ \hline
		$\sigma = 20$ $s=0.3$ $r=0.3$ & 22.85/60.65 & 24.33/69.08 & \textbf{25.06/72.42} \\ \hline
		$\sigma = 30$ $s=0.3$ $r=0.4$ & 20.26/46.65 & 21.16/54.76 & \textbf{22.40/62.59} \\ \hline \hline
	\end{tabular}
\end{table}

\begin{table}[h]
	\centering
	\setlength{\abovecaptionskip}{0.cm}
	\setlength{\belowcaptionskip}{0.cm}
	\fontsize{6.5pt}{\baselineskip}\selectfont
	\caption{The average PSNR(dB)/SSIM(\%) results of mixed noise removal on BSD100 dataset.}
	\label{Exp:BSD}
	\begin{tabular}{cccccc}
		\hline \hline
		\multicolumn{2}{c}{Mixed Noise}           & MBM3D \cite{dabov2007image}        & WESNR \cite{jiang2014mixed}       & WJSR \cite{liu2017weighted}        & RWE (ours)                  \\ \hline
		\multirow{3}{*}{$\sigma = 10$} & $s = 30\%$ & 28.00/82.38 & 28.16/80.87 & 28.92/83.70 & \textbf{29.61/86.00} \\ 
		& $s = 50\%$ & 25.79/76.58 & 26.57/77.17 & 27.50/79.04 & \textbf{27.94/81.89} \\ 
		& $s = 70\%$ & 23.28/66.24 & 24.36/70.33 & 25.33/70.25 & \textbf{25.54/71.60} \\ \hline
		\multirow{3}{*}{$\sigma = 30$} & $s = 30\%$ & 25.12/65.85 & 24.34/66.45 & 25.39/67.65 & \textbf{25.80/69.81} \\ 
		& $s = 50\%$ & 23.56/58.69 & 22.88/62.68 & 24.30/62.82 & \textbf{24.82/66.27} \\ 
		& $s = 70\%$ & 21.40/47.90 & 20.74/55.24 & 22.45/55.93 & \textbf{23.16/60.28} \\ \hline
		\multirow{3}{*}{$\sigma = 50$} & $s = 30\%$ & 23.28/55.76 & 17.28/36.95 & 23.58/60.37 & \textbf{24.00/60.66} \\
\hline \hline
	\end{tabular}
\end{table}

\begin{figure*}[h] \vspace*{-.5cm}
	\setcounter{subfigure}{0} 
	\centering
	\subfloat{
		\begin{minipage}{0.19\textwidth}
			\centering
			\setlength{\abovecaptionskip}{-0.1cm}
			\setlength{\belowcaptionskip}{-0.1cm}
			\includegraphics[width=1\columnwidth]{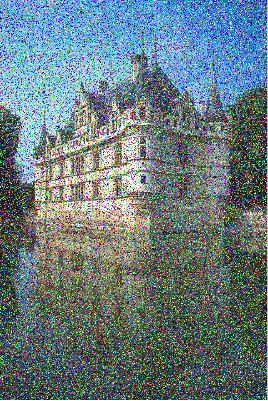}
			\label{102061:noisy}
			\caption*{(a) Noisy image (PSNR)}
		\end{minipage}
	}
	\subfloat{
		\begin{minipage}{0.19\textwidth}
			\centering
			\setlength{\abovecaptionskip}{-0.1cm}
			\setlength{\belowcaptionskip}{-0.1cm}
			\includegraphics[width=1\columnwidth]{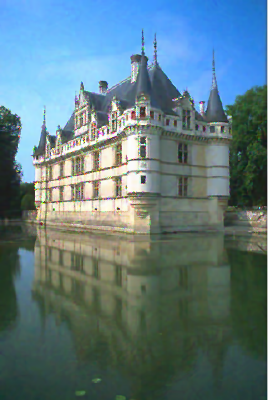}
			\label{102061:BM3D}
			\caption*{(b) MBM3D \cite{dabov2007image} (27.05 dB)}
		\end{minipage}
	}  	
	\subfloat{
		\begin{minipage}{0.19\textwidth}
			\centering
			\setlength{\abovecaptionskip}{-0.1cm}
			\setlength{\belowcaptionskip}{-0.1cm}
			\includegraphics[width=1\columnwidth]{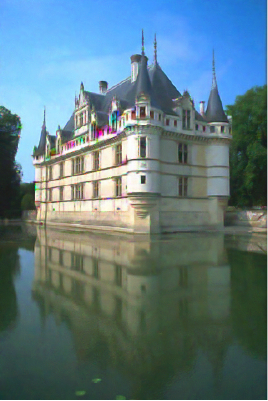}
			\label{102061:WESNR}
			\caption*{(c) WESNR \cite{jiang2014mixed} (27.15 dB)}
		\end{minipage}
	}  	 
	\subfloat{
	\begin{minipage}{0.19\textwidth}
		\centering
		\setlength{\abovecaptionskip}{-0.1cm}
		\setlength{\belowcaptionskip}{-0.1cm}
		\includegraphics[width=1\columnwidth]{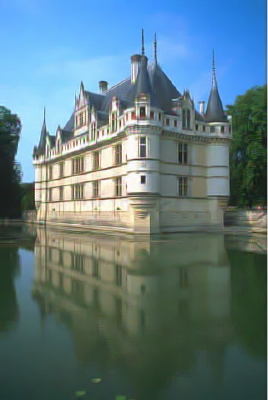}
		\label{102061:WJSR}
		\caption*{(d) WJSR \cite{liu2017weighted} (28.95 dB)}
	\end{minipage}
}  	
	\subfloat{
		\begin{minipage}{0.19\textwidth}
			\centering
			\setlength{\abovecaptionskip}{-0.1cm}
			\setlength{\belowcaptionskip}{-0.1cm}
			\includegraphics[width=1\columnwidth]{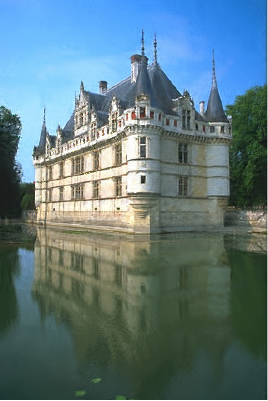}
			\label{102061:Proposed}
			\caption*{(e) Our RWE (\textbf{29.33} dB)}
		\end{minipage}
	}  	
	\caption{Denoising on color image \textit{102061} from BSD100 dataset corrupted with AWGN+SPIN ($\sigma = 10$, $s = 30\%$).}
	\label{Exp:102061}
\end{figure*}

\begin{figure*}[ht] \vspace*{-.5cm} 
	\setcounter{subfigure}{0} 
	\centering
	\subfloat{
		\begin{minipage}{0.19\textwidth}
			\centering
			\setlength{\abovecaptionskip}{-0.1cm}
			\setlength{\belowcaptionskip}{-0.1cm}
			\includegraphics[width=1\columnwidth]{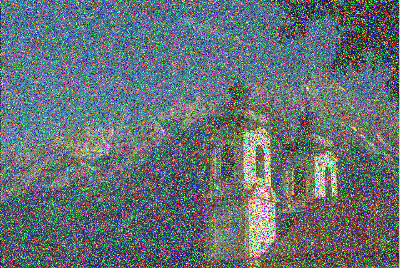}
			\label{126007:noisy}
			\caption*{(a) Noisy image (PSNR)}
		\end{minipage}
	}
	\subfloat{
		\begin{minipage}{0.19\textwidth}
			\centering
			\setlength{\abovecaptionskip}{-0.1cm}
			\setlength{\belowcaptionskip}{-0.1cm}
			\includegraphics[width=1\columnwidth]{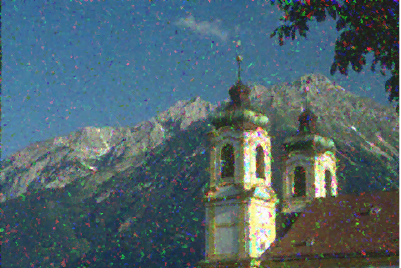}
			\label{126007:BM3D}
			\caption*{(b) MBM3D \cite{dabov2007image} (22.68 dB)}
		\end{minipage}
	}  	
	\subfloat{
		\begin{minipage}{0.19\textwidth}
			\centering
			\setlength{\abovecaptionskip}{-0.1cm}
			\setlength{\belowcaptionskip}{-0.1cm}
			\includegraphics[width=1\columnwidth]{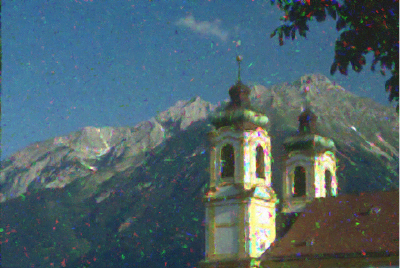}
			\label{126007:WESNR}
			\caption*{(c) WESNR \cite{jiang2014mixed} (24.30 dB)}
		\end{minipage}
	}  	 
	\subfloat{
		\begin{minipage}{0.19\textwidth}
			\centering
			\setlength{\abovecaptionskip}{-0.1cm}
			\setlength{\belowcaptionskip}{-0.1cm}
			\includegraphics[width=1\columnwidth]{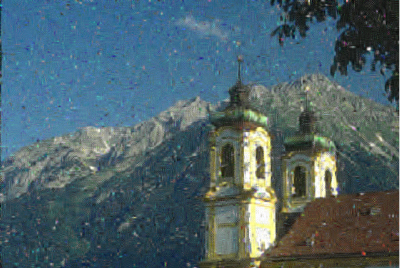}
			\label{126007:WJSR}
			\caption*{(d) WJSR \cite{liu2018mixed} (24.14 dB)}
		\end{minipage}
	}  	
	\subfloat{
		\begin{minipage}{0.19\textwidth}
			\centering
			\setlength{\abovecaptionskip}{-0.1cm}
			\setlength{\belowcaptionskip}{-0.1cm}
			\includegraphics[width=1\columnwidth]{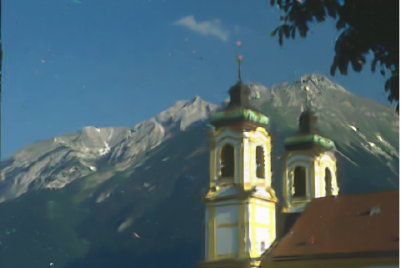}
			\label{126007:Proposed}
			\caption*{(e) Our RWE (\textbf{26.40} dB)}
		\end{minipage}
	}  	
		\caption{Denoising on color image \textit{126007} from BSD100 dataset corrupted with AWGN+RVIN ($\sigma = 10$, $r = 50\%$).}
	\label{Exp:126007}
\end{figure*}

\begin{figure*}[h] \vspace*{-.5cm} 
	\setcounter{subfigure}{0} 
	\centering
	\subfloat{
		\begin{minipage}{0.19\textwidth}
			\centering
			\setlength{\abovecaptionskip}{-0.1cm}
			\setlength{\belowcaptionskip}{-0.1cm}
			\includegraphics[width=1\columnwidth]{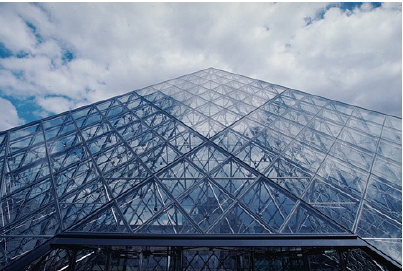}
			\label{color:original}
			\caption*{(a) Original image}
		\end{minipage}
	}  	
	\subfloat{
		\begin{minipage}{0.19\textwidth}
			\centering
			\setlength{\abovecaptionskip}{-0.1cm}
			\setlength{\belowcaptionskip}{-0.1cm}
			\includegraphics[width=1\columnwidth]{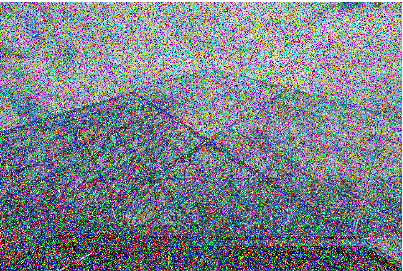}
			\label{color:noisy}
			\caption*{(b) Noisy image (PSNR (dB))}
		\end{minipage}
	}
	\subfloat{
		\begin{minipage}{0.19\textwidth}
			\centering
			\setlength{\abovecaptionskip}{-0.1cm}
			\setlength{\belowcaptionskip}{-0.1cm}
			\includegraphics[width=1\columnwidth]{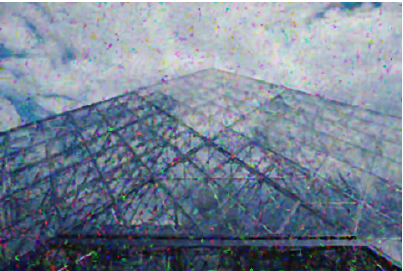}
			\label{color:BM3D}
			\caption*{(c) MBM3D \cite{dabov2007image} (22.11 dB)}
		\end{minipage}
	}  	
	\subfloat{
		\begin{minipage}{0.19\textwidth}
			\centering
			\setlength{\abovecaptionskip}{-0.1cm}
			\setlength{\belowcaptionskip}{-0.1cm}
			\includegraphics[width=1\columnwidth]{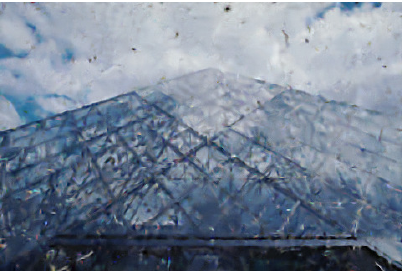}
			\label{color:WESNR}
			\caption*{(d) WESNR \cite{jiang2014mixed} (23.09 dB)}
		\end{minipage}
	}  	 
	\subfloat{
		\begin{minipage}{0.19\textwidth}
			\centering
			\setlength{\abovecaptionskip}{-0.1cm}
			\setlength{\belowcaptionskip}{-0.1cm}
			\includegraphics[width=1\columnwidth]{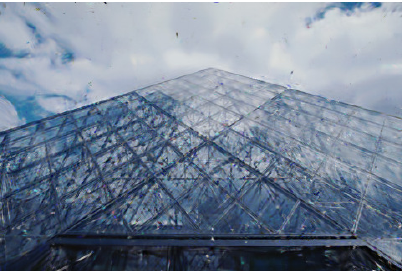}
			\label{color:Proposed}
			\caption*{(e) Our RWE (\textbf{23.63} dB)}
		\end{minipage}
	}  	
	\caption{Denoising on color image \textit{223061} from BSD100 corrupted with AWGN+SPIN+RVIN ($\sigma = 10$, $s = 20\%$, $r= 30\%$). }
	\label{Exp:color}
\end{figure*}

\begin{table*}[h]
	\setlength{\abovecaptionskip}{0.cm}
	\setlength{\belowcaptionskip}{-0.2cm}
	\centering
	\caption{Denoising results (PSNR(dB)/SSIM(\%)) of different methods on Barbara image with AWGN+RVIN. }
	\label{exp:deepbarbara}
	\begin{tabular}{cccccc} 
		\hline\hline
		& \multicolumn{2}{c}{$\sigma$ = 10 }                                &                      & \multicolumn{2}{c}{$\sigma$ = 15 }                                 \\ 
		\cline{2-3}\cline{5-6}
		& r = 0.2                           & r = 0.3                           &                      & r =  0.2                           & r = 0.3                            \\ 
		\hline
		\multicolumn{1}{c}{DnCNN \cite{zhang2017beyond}}  & \multicolumn{1}{c}{~5.67/00.35}  & \multicolumn{1}{c}{~5.58/00.33}  & \multicolumn{1}{c}{} & \multicolumn{1}{c}{~5.75/00.43}  & \multicolumn{1}{c}{~5.72/00.41}   \\
		\multicolumn{1}{c}{FOCNet \cite{jia2019focnet}} & \multicolumn{1}{c}{16.54/28.27} & \multicolumn{1}{c}{14.64/20.06} & \multicolumn{1}{c}{} & \multicolumn{1}{c}{16.35/27.17} & \multicolumn{1}{c}{14.51/19.46}  \\EM-CNN \cite{wang2019variational}                    & \textbf{29.55}/87.49            & 27.41/83.13                     &                      & \textbf{28.09}/82.22            & 25.73/77.50                      \\
		RWE (ours)                 &28.94/\textbf{90.00}&\textbf{27.57}/\textbf{87.27}&                      &27.45/\textbf{82.90}&\textbf{26.70}/\textbf{82.17}\\
		\hline\hline
	\end{tabular}
\end{table*}

\begin{table*}[h]
	\setlength{\abovecaptionskip}{0.cm}
	\setlength{\belowcaptionskip}{-0.2cm}
	\footnotesize
	\centering
	\caption{\footnotesize The average PSNR(dB)/SSIM comparison for single type noise removal on 12 standard test images.}
	\label{Exp:only}
	\begin{tabular}{cccc|ccc|ccc} 
		\hline \hline
		\multirow{2}{*}{} & \multicolumn{3}{c|}{Only SPIN}         & \multicolumn{3}{c|}{Only RVIN}          & \multicolumn{3}{c}{Only AWGN}            \\
		& s = 20\%    & s = 30\%    & s = 40\%    & r = 20\%    & r = 30\%    & r = 40\%    & $\sigma = 10$          & $\sigma = 20 $         & $\sigma = 30$          \\ 
		\cline{2-10}
		MBM3D \cite{dabov2007image}             & 31.10/93.37 & 29.47/91.24 & 28.01/88.37 & 29.21/88.14 & 26.67/79.67 & 22.35/67.56 & 31.32/87.82 & 28.81/81.55 & 27.09/75.93  \\
		WESNR \cite{jiang2014mixed}          & 31.17/88.88 & 30.60/88.19 & 29.91/87.13 & 30.84/89.56 & 29.19/86.93 & 27.21/82.08 & 30.89/86.87 & 28.41/79.65 & 27.00/75.59  \\
		WJSR  \cite{liu2017weighted}            & 12.30/17.83 & 9.76/13.17  & 9.30/10.49  & 13.75/22.20 & 12.11/16.34 & 10.97/12.80 & 32.18/89.86 & 29.47/83.85 & 27.86/79.02  \\
		RWE (ours)        & \textbf{34.21}/\textbf{93.82}& \textbf{34.03}/\textbf{93.15} & \textbf{30.11}/\textbf{89.71} & \textbf{31.09}/\textbf{90.17} & \textbf{29.36}/\textbf{86.89} & \textbf{27.63}/\textbf{81.08} & \textbf{32.25}/\textbf{89.91} & \textbf{29.64}/\textbf{83.89} & \textbf{28.02}/\textbf{79.04}   \\
		\hline \hline
	\end{tabular}
\end{table*}

\section{Experimental Results} \label{Sec:Experiment}
In this section, experimental results are presented to verify our proposed RWE method for mixed noise removal. Both PSNR (Peak Signal to Noise Ratio) and SSIM (Structural Similarity Image Measurement) \cite{wang2004image} are calculated to evaluate the quality of the denoised images.


\subsection{Parameter setting} \label{Exp:paramets}
In our experiments, the exemplar image patches are extracted in every 4 pixels along both horizontal and vertical directions, which have been used in \cite{dong2013nonlocal,dong2015image}. There are two method parameters need to be tuned in our method, \ie, patch size $\sqrt{m} \times \sqrt{m}$ and the number of similar patches $K$.

To investigate the sensitivity of these two parameters, experiments with respect to $\sqrt{m} \times \sqrt{m}$ and $K$ for mixed noise removal are conducted, respectively. The performance comparison on test image \textit{Barbara} with various $\sqrt{m} \times \sqrt{m}$ and $K$ are presented in Fig.~\ref{parametersetting}. We can see that the performance of our RWE method is not sensitive to $\sqrt{m} \times \sqrt{m}$ and $K$. As $\sqrt{m} = 11$ and $K=60$ generally leads to the best performance, we choose $\sqrt{m} = 11$ and $K = 60$ empirically in our experiment.


\subsection{The Effectiveness of Pareto Prior}\label{Exp:effective}

\subsubsection{\textbf{Ablation Study}}
To validate the effectiveness of our proposed Pareto prior, we conduct ablation study
on four variants of our proposed method: we update $\bm{W}$ in our method (a) with detection results (without Pareto prior, like WJSR \cite{liu2017weighted}); (b) with WESNR's method \cite{jiang2014mixed}; (c) with RCSR's prior \cite{liu2018mixed}; (d) with our Pareto prior. We conduct experiments on two gray-scale images \textit{Lena} and \textit{Barbara} with various mixed noise and the corresponding parameters are set as the recommended values by their authors. The PSNR comparisons are illustrated in Fig.~\ref{effectiveness}.  We can see that our proposed Pareto prior achieve larger value of PSNR than others under different levels of mixed noise. The results validate that our Pareto prior can successfully improve the performance.

\subsubsection{\textbf{The Accuracy and Robustness}}

The accuracy and robustness of the weighting matrix estimator is evaluated according to the mean and variance of the weighted residuals between the noisy observations $\bm{Y}$ and the estimated clean images $\bm{X}$. Apparently, the ground truth of the mean and variance is just corresponding to the true AWGN. Thus, with more accurate weighting matrix $\bm{W}$, the distribution of the weighted residuals will be closer to the true AWGN. We conduct experiments to estimate the means and variances of the weighted residuals by using different methods. All estimated results are presented in Table~\ref{tab:robust}. One can see that the variance estimated by WESNR tends to be larger than the ground truth, which lead to the under-weighting problem. Thus the effects of IN cannot be suppress effectively by WESNR. On the other hand, RCSR usually produces the smaller variance than the ground truth, which lead to the over-weighting problem. Thus RCSR tends to over-smooth the image details. Visual results in Fig.~\ref{fig:highlight} support our findings. We can see that the mean and variance estimated by our RWE method are closer to the ground truth. This is consistent with our observation in Fig.~\ref{fig:highlight}(g). Therefore, it can verify that our RWE method can estimate weighting matrix more accurately and robustly for different levels of mixed noise.

\subsection{Comparison with State-of-the-arts}\label{Exp:results}

We perform our comparison experiments on two commonly used standard image datasets: \textit{12 classic standard test images} and \textit{Berkeley Segmentation Dataset} (BSD100) \cite{MartinFTM01}. Three types of mixed noise, \ie, AWGN+SPIN, AWGN+RVIN, AWGN+SPIN+RVIN are considered in experiments. To generate different mixed noise levels, the standard deviations $\sigma$ of AWGN  are varied from 10 to 50, the SPIN ratio $s$ from 20\% to 70\%, and the RVIN ratio $r$ from 20\% to 70\%. For the sake of fairness, we utilize AMF  \cite{hwang1995adaptive} (for SPIN ) and ACWMF \cite{chen2001adaptive} (for RVIN) as the initialization of our algorithm, which are commonly used in most mixed noise removal methods\cite{liu2017weighted,jiang2014mixed,liu2018mixed,xiao2011restoration,jiang2015mixed,huang2017mixed}.

\subsubsection{\textbf{Comparison with Traditional Methods}}

we first compare our proposed RWE to other representative methods for mixed noise removal task, i.e. WESNR \cite{jiang2014mixed} and WJSR \cite{liu2017weighted}. Although the codes of RCSR \cite{liu2018mixed} is not released,  WESNR is instead compared since it is reported in \cite{liu2018mixed} that RCSR is slightly outperforms WESNR. Furthermore, BM3D \cite{dabov2007image} is also exploited as the reference even though it is not designated for mixed noise removal task. To deal with mixed noises, BM3D is coupled with the median filter, i.e. AMF and ACWMF, and named as MBM3D. The codes of all the competing methods are provided by the original paper with default parameter settings.



\begin{figure*}[h]
	\setcounter{subfigure}{0} 
	\centering
	\subfloat{
		\begin{minipage}{0.19\textwidth}
			\centering
			\setlength{\abovecaptionskip}{-0.1cm}
			\setlength{\belowcaptionskip}{-0.1cm}
			\includegraphics[width=1\columnwidth]{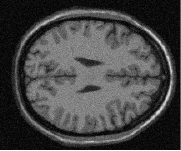}
			\label{MRI:original}
			\caption*{(a) Real noisy image}
		\end{minipage}
	}  	
	\subfloat{
		\begin{minipage}{0.19\textwidth}
			\centering
			\setlength{\abovecaptionskip}{-0.1cm}
			\setlength{\belowcaptionskip}{-0.1cm}
			\includegraphics[width=1\columnwidth]{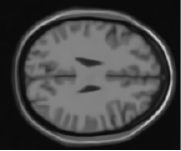}
			\label{MRI:BM3D}
			\caption*{(b) MBM3D \cite{dabov2007image}}
		\end{minipage}
	}
	\subfloat{
		\begin{minipage}{0.19\textwidth}
			\centering
			\setlength{\abovecaptionskip}{-0.1cm}
			\setlength{\belowcaptionskip}{-0.1cm}
			\includegraphics[width=1\columnwidth]{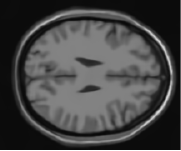}
			\label{MRI:WESNR}
			\caption*{(c) WESNR \cite{jiang2014mixed}}
		\end{minipage}
	}  	
	\subfloat{
		\begin{minipage}{0.19\textwidth}
			\centering
			\setlength{\abovecaptionskip}{-0.1cm}
			\setlength{\belowcaptionskip}{-0.1cm}
			\includegraphics[width=1\columnwidth]{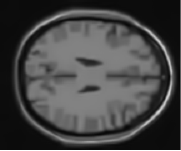}
			\label{MRI:WJSR}
			\caption*{(d) WJSR \cite{liu2018mixed}}
		\end{minipage}
	}  	 
	\subfloat{
		\begin{minipage}{0.19\textwidth}
			\centering
			\setlength{\abovecaptionskip}{-0.1cm}
			\setlength{\belowcaptionskip}{-0.1cm}
			\includegraphics[width=1\columnwidth]{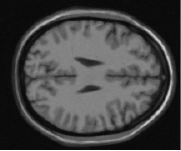}
			\label{MRI:Proposed}
			\caption*{(e) Our RWE}
		\end{minipage}
	}  	
	\caption{  Denoising on realistic image.}
	\label{Exp:MRI}
\end{figure*}

\textbf{Results on Gray-scale Images.} The PSNR and SSIM results on 12 classic standard test images are presented in Table~\ref{Exp:SPIN} (for AWGN+SPIN removal), Table~\ref{Exp:RVIN} (for AWGN+RVIN removal) and Table~\ref{Exp:SPRV} \footnotemark[2] (for AWGN+SPIN+RVIN removal), respectively. The average PSNR/SSIM values on BSD100 for three types of mixed noise are presented in Table~\ref{Exp:BSD}. One can see that the performance of WESNR and WJSR decreases rapidly with increasing mixed noise levels, and our proposed RWE outperforms other competing methods with various mixed noise levels. The PSNR/SSIM gains over other competing methods can be more than 1dB/6\% on a parts of images. Some visual comparisons of denoised images on the three types of mixed noises are presented in Fig.~\ref{Exp:lena}, Fig.~\ref{Exp:barbara} and Fig.~\ref{Exp:peppers}. All these verify the superiority of our proposed RWE for mixed noise removal. 


\footnotetext[2]{For the mixed noise of AWGN+SPIN+RVIN, we do not compare with WJSR \cite{liu2017weighted}, as it cannot deal with AWGN+SPIN+RVIN.}

\textbf{Results on Color Images.} We also conduct experiments on color images from BSD 100, where all competing methods are applied independently on each color channel. Fig.~\ref{Exp:102061}, Fig.~\ref{Exp:126007} and Fig.~\ref{Exp:color} present the visual results of color images from BSD100. Our RWE method consistently achieve the best visual performance in terms of effectively removing mixed noise while preserving image details.




\subsubsection{\textbf{Comparison with Deep Learning based  Methods}} 

Currently, very few studies focus on deep learning based mixed noise removal \cite{wang2019variational,ding2019improved}. In order to better evaluate the performance, we compare our proposed RWE to some state-of-the-art deep learning based methods, including DnCNN \cite{zhang2017beyond}, FOCNet \cite{jia2019focnet}, and EM-CNN \cite{wang2019variational}. DnCNN \cite{zhang2017beyond} and FOCNet \cite{jia2019focnet} are the recent deep learning based denoising methods for generalized blind image denoisng problem. EM-CNN is designated for AWGN+RVIN removal task, thus we conduct the evaluation on Barbara contaminated by AWGN+RVIN, as shown in Table~\ref{exp:deepbarbara}. The performance of both DnCNN and FOCNet are largely degraded due to the existence of RVIN. Our proposed RWE and EM-CNN have comparable performance in terms of PSNR where RWE outperforms EM-CNN for the case of high level of RVIN ratio. On the other hand, RWE can achieve the highest value of SSIM in the competing algorithms. 

Even though, denoising based on deep learning techniques emerging recently often outperforms most of the traditional methods, it is largely dependent on the training data and often requires a long-time training stage. While according to the experimental results, our proposed RWE method can achieve the better performance than the deep learning based methods without training stage. Furthermore, the proposed RWE can address three types of mixed noises simultaneously, i.e. SPIN+AWGN, RVIN+AWGN and SPIN+RVIN+AWGN, which have not been considered in deep learning based methods.


\subsubsection{\textbf{Comparison of Single Type Noise Removal}} We conduct experiments to demonstrate the efficiency of our RWE method for single type noise removal. The average PSNR/SSIM values for singe type noise removal are presented in Table~\ref{Exp:only}. We can see that WJSR degrades largely for single SPIN or RVIN removal case. This is mainly due to the fact that it is quite difficult to select some appropriate parameters empirically for dictionary learning and sparse coding under different noise environment. In contrast, our RWE can update shape parameter adaptively. Hence our RWE have higher average PSNR and SSIM than other methods. 


\subsubsection{\textbf{Comparison on Realistic Images}}
To show the effectiveness of our RWE method, we conduct experiments on some real noise images \cite{zhuang2016mixed,liu2013weighted}. The comparison results are illustrated in Fig~\ref{Exp:MRI}. One can see that WJSR over-smooths some detail structures and our RWE method achieves the better visual quality than other methods.

\begin{figure}[h]
	\setcounter{subfigure}{0} 
	\centering
	\subfloat{
		\begin{minipage}{0.245\textwidth}
			\centering
			\setlength{\abovecaptionskip}{-0.1cm}
			\setlength{\belowcaptionskip}{-0.1cm}
			\includegraphics[width=1\columnwidth]{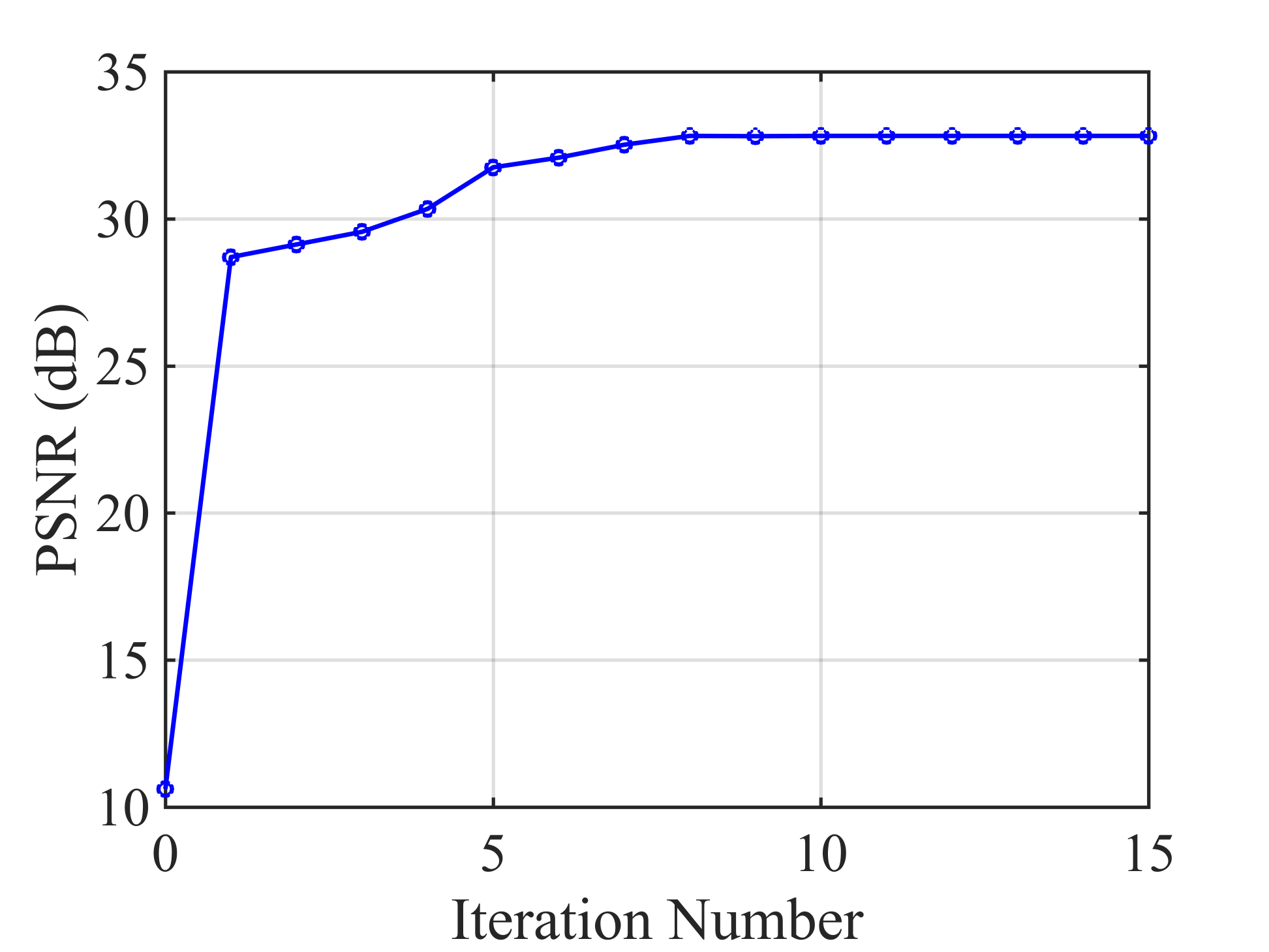}
			\label{convergence:inner}
			\caption*{(a)}
		\end{minipage}
	}
	\subfloat{
		\begin{minipage}{0.245\textwidth}
			\centering
			\setlength{\abovecaptionskip}{-0.1cm}
			\setlength{\belowcaptionskip}{-0.1cm}
			\includegraphics[width=1\columnwidth]{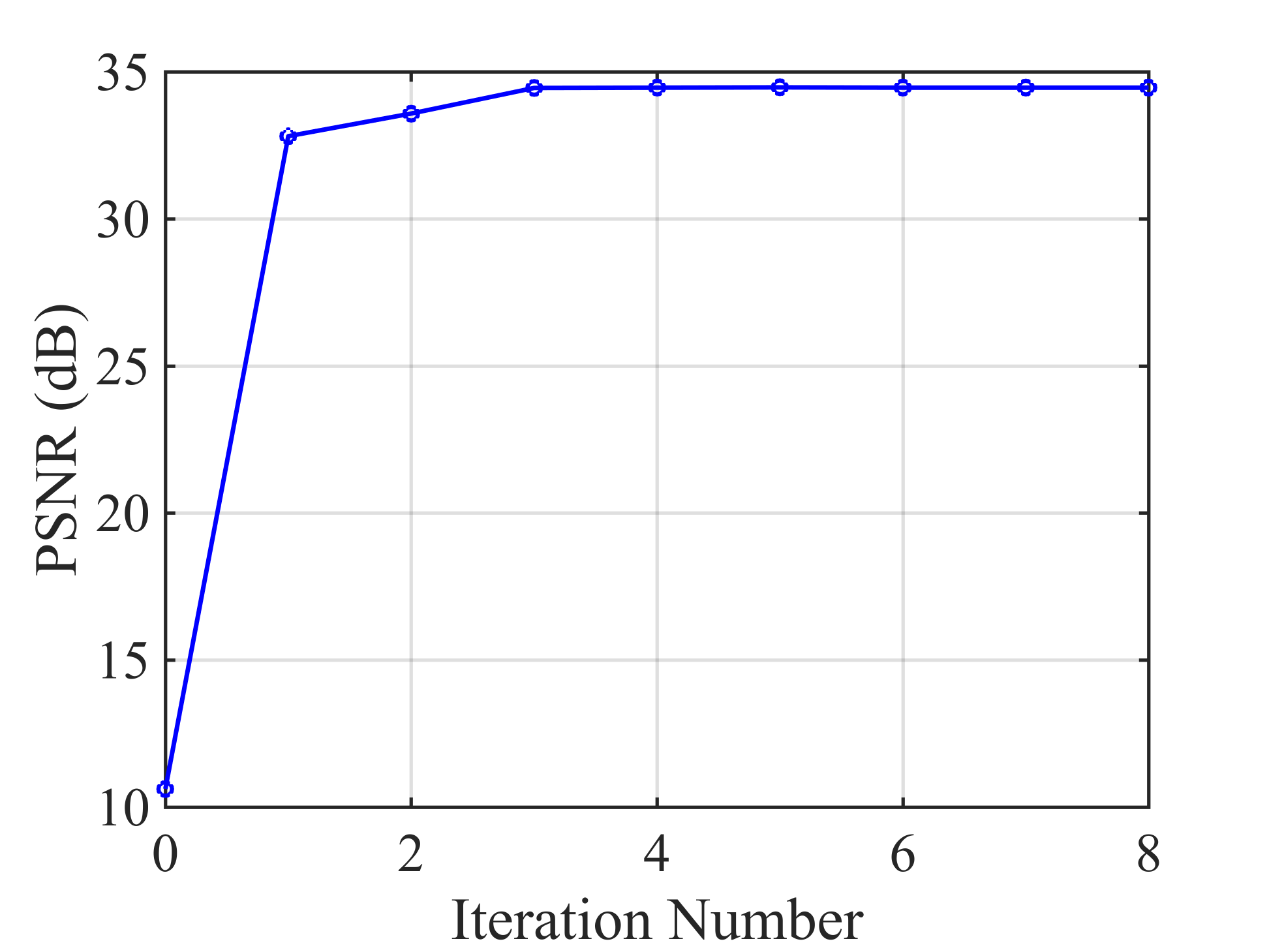}
			\label{convergence:outer}
			\caption*{(b)}
		\end{minipage}
	}  	
	\caption{The PSNR curves with mixed noise (AWGN+SPIN $\sigma =10 $ $s = 0.3$) on test $Lena$ image:  (a) the curve of PSNR versus iteration number for solving the $\bm{X}$ subproblem,  and (b) the curve of PSNR versus iteration number for outer loop. }
	\label{Exp:convergence}
\end{figure}

\begin{table}[h]
\fontsize{6.5pt}{\baselineskip}\selectfont
	\centering
	\caption{The  average  computational time (s) of different methods to process a  $512 \times 512$ image}
	\label{Exp:computationaltime}
	\begin{tabular}{ccccc}
		\hline \hline
		Method           & MBM3D \cite{dabov2007image}         & WESNR \cite{jiang2014mixed} & WJSR \cite{liu2017weighted} & RWE (ours)  \\ \hline
		Average Time (s) & 2.5 (C++)& 69.4 (Matlab)&   642.5 (Matlab) & 118.1 (Matlab)\\ \hline \hline
	\end{tabular}
\end{table}


\subsection{Convergence and Computational Time }\label{Exp:time}
Now we carry out experiment to verify the convergence of our algorithm, which has been discussed in Section~\ref{Sec:convergence}. As presented in Fig.~\ref{convergence:inner}, the inner loop of solving the $\bm{X}$ subproblem converges after 8 iterations. Thus we usually set the iteration number of inner loop $C = 8$. For the outer loop, Fig.~\ref{convergence:outer} shows that our proposed RWE algorithm converges quickly in about 3 iterations. And thus we set the maximum iteration number $I_{max} = 3$. 

Furthermore, we compare the average computational time of all competing methods by processing a $512 \times 512$ \textit{Lena} image. The results are presented in Table~\ref{Exp:computationaltime}. All experiments are implemented on Matlab 2015b with Intel Core i7-6900K 3.7G Hz CPU and 32GB RAM. One can see that BM3D is the fastest and it spends only 2.5s. This is due to the fact that BM3D is implemented with compiled C++ mex-function and further sped up by optimization and parallelization. For other competing algorithms including our proposed RWE, they are implemented purely in Matlab, less optimized and with single thread. Noted that nuclear norm minimization based low-rank approximation in our RWE is of high computational complexity. It can be further sped up by using parallel computational technique, just like BM3D. 

We also show the computational complexity here. For each iteration, the computational complexity of our proposed algorithm mainly consists of three parts: 1) solving the $\bm{W}$ subproblem; 2) solving the $\bm{X}$ subproblem; 3) solving the $\sigma$ subproblem. The complexity of solving $\bm{W}$ subproblem is $O(MN)$. And the complexity of solving $\bm{X}$ subproblem can be divided into two parts: the k-NN patch grouping and singular value thresholding. The complexity of computing the k-NN patch grouping is $O(MNKs^2m/16)$, where $K$ is the number of similar patches in each group, $s$ is the size of search window $s \times s$, $m$ is the number of pixels in each group. Then the complexity of computing singular value thresholding is $O(MNmK^2/16)$. The computational complexity of solving $\sigma$ subproblem is $O(MN)$. Therefore, the overall complexity of one iteration is $O(MNKs^2m/16)$.

\section{Conclusion} \label{Sec:Conclusion}

For the task of mixed noise removal, one can boost the denoising performance by compensating the effects of IN with a weighting matrix. Accurately estimating the weighting matrix plays an important role. According to the observation of few IN and many AWGN, we propose to exploit the Pareto prior to describe the behavior of the weighting matrix, leading to an accurate and robust weighting matrix estimator. Together with the nonlocal low rank approximation, a mixed noise removal method with robust weighting matrix estimation (RWE) is obtained. We further develop an alternative optimization algorithm to solve our proposed model. Experimental results on widely used image datasets demonstrate the effectiveness and superiority of our proposed method. 

In future work, we will consider to learn the structure relationship of weights between different image regions from the external dataset using deep learning based algorithm. Then we can integrate this learned structure prior into our variational model to estimate the weights more precisely.

\ifCLASSOPTIONcaptionsoff
  \newpage
\fi

\end{document}